\documentstyle[pra,aps,eqsecnum,epsf,multicol]{revtex}

\begin{document}

\title{Random antiferromagnetic quantum spin chains:\\
Exact results from scaling of rare regions}

\author{Ferenc Igl\'oi$^{1,2}$, R\'obert Juh\'asz$^{2,1}$ and Heiko Rieger$^{3}$}

\address{
$^1$ Research Institute for Solid State Physics and Optics, 
H-1525 Budapest, P.O.Box 49, Hungary\\
$^2$ Institute for Theoretical Physics,
Szeged University, H-6720 Szeged, Hungary\\
$^3$ FB 10.1 Theoretische Physik, Universit\"at des Saarlandes, 
     66041 Saarbr\"ucken, Germany\\
}

\date{April 11, 1999}

\maketitle

\begin{abstract}
We study  $XY$ and dimerized $XX$ spin-$1/2$ chains with
random exchange couplings by analytical and numerical methods and
scaling considerations. We extend previous investigations to dynamical
properties, to surface quantities and operator profiles, and give a
detailed analysis of the Griffiths phase. We present a phenomenological scaling
theory of average quantities based on the scaling properties of rare
regions, in which the distribution of the couplings follows a surviving random
walk character. Using this theory we have
obtained the complete set of critical decay exponents of the random $XY$ and
$XX$ models, both in the volume and at the surface. The scaling results
are confronted with numerical calculations based on a mapping to free
fermions, which then lead to an exact correspondence with directed walks. The
numerically calculated critical operator profiles on large finite systems
$(L \le 512)$ are found to follow conformal predictions with the decay
exponents of the phenomenological scaling theory.  Dynamical correlations in
the critical state are in average logarithmically slow and their distribution
show multi-scaling character. In the Griffiths phase, which is an extended part
of the off-critical region
average autocorrelations have a power-law form with
a non-universal decay exponent, which is analytically calculated. We note on
extensions of our work to the random antiferromagnetic $XXZ$ chain and to
higher dimensions.
\end{abstract}


\newcommand{\bc}{\begin{center}}
\newcommand{\ec}{\end{center}}
\newcommand{\be}{\begin{equation}}
\newcommand{\ee}{\end{equation}}
\newcommand{\ba}{\begin{array}}
\newcommand{\ea}{\end{array}}
\newcommand{\beqn}{\begin{eqnarray}}
\newcommand{\eeqn}{\end{eqnarray}}

\begin{multicols}{2}
\narrowtext

\section{Introduction}
Quantum spin chains exhibit many interesting physical properties at
low temperatures which are related to the behavior of their ground
state and low-lying excitations. In this context one should mention
quasi-long-range-order (QLRO), topological order and quantum phase
transitions, which have purely quantum mechanical origin. Considering
isotropic antiferromagnetic chains for integer spins there is a gap,
whereas half integer spin chains are gap-less\cite{haldane}. However,
alternating couplings in spin-$1/2$ chains yield a dimerized ground
state that has physical properties similar to the spin-$1$
chain: there is a finite gap, spatial correlations decay exponentially
and there is string topological order.

Randomness may have a profound effect on the physical properties of
quantum spin chains, as demonstrated by recent analytical and
numerical studies\cite{qsg}. As an interplay of randomness and quantum
fluctuations there are new exotic phases in disordered quantum spin
chains, which are not present in classical random or pure quantum
systems. It has been noticed, that pure gap-less systems are generally
unstable against weak randomness\cite{ma,fisherxx}, whereas for gaped
systems a finite amount of disorder is necessary to destroy the
gap\cite{hyman,fabrizio,monthus} (but see also\cite{hida}).

Among the theoretical methods developed for disordered quantum spin
chains one very powerful procedure is the renormalization group (RG)
approach introduced by Dasgupta and Ma\cite{ma}. This RG method, which
is expected to be asymptotically exact at large scales, i.e. close to
critical points, has been applied for a number of random quantum
systems. The fixed point distribution of the RG transformation has
been obtained analytically for some random quantum spin chains, among
others for the transverse Ising spin chain\cite{fisher}, the
spin-$1/2$ Heisenberg and related spin chains with random
antiferromagnetic couplings\cite{fisherxx}. On the other hand some
other one-dimensional problems ($S=1/2$ Heisenberg chain with mixed
ferromagnetic and antiferromagnetic
couplings\cite{westerberg,hikihara}, $S=1$ antiferromagnetic chain
with and without biquadratic exchange\cite{hyman,fabrizio,monthus},
etc.), as well as higher dimensional random quantum systems\cite{RG}
have been studied by numerical implementation of the RG procedure.
Comparing the RG results with those obtained by direct numerical
evaluation of the singular
quantities\cite{youngrieger,profiles,bigpaper,young} and by other
exact\cite{mckenzie,bigpaper} and numerical methods one has obtained a
good agreement in the vicinity of the critical point.

There are, however, other interesting singular quantities, which are
not accessible by the RG method. We mention, among others, the
dynamical correlations\cite{riegerigloi} and the behavior of the
system far away from the critical point in the Griffiths
phase\cite{griffiths}, which denotes an extended region of the
parameter space around the critical point. In the Griffiths phase the
system is gap-less, thus dynamical correlations decay with a power-law,
however there is long-range-order with exponentially decaying spatial
correlations.  For the random quantum Ising chain dynamical
correlations, both at the critical point and in the Griffiths phase
have been exactly determined\cite{diffusion,itr,ijr} using a
mapping to the Sinai model\cite{sinai}, i.e. random walk in a random
environment.

In this paper we are going to study the $S=1/2$ disordered $XX$ and
$XY$ spin chains by analytical and numerical methods and by
phenomenological scaling theory. The RG treatment of the problem by
Fisher\cite{fisherxx} predicts the antiferromagnetic random $XX$ fixed
point to control the critical behavior of the antiferromagnetic
Heisenberg ($XXX$) model, too. Furthermore, for random isotropic
chains the RG approach predicts QLRO, thus the average spatial
correlations of different components of the spin decay with a
power-law. In this so called random singlet (RS) phase all spins are
paired and form singlets, however, the distance between the two spins
in a singlet pair can be arbitrarily large. Then these weakly coupled
singlets dominate the average correlation function, therefore all
components of the correlation function are predicted to decay with the
same exponent.

Leaving the critical state by introducing either anisotropy or
dimerization, randomness will drive the system into the
Griffiths phase. As shown by an RG
analysis\cite{hyman}, applicable in the vicinity of the RS fixed
point, the Griffiths phase is characterized by the dynamical
exponent, $z$, defined by the asymptotic relation between relevant
time ($t_r$) and length scales ($\xi$) as
\be
t_r \sim \xi^{z}\;.
\label{z}
\ee
The dynamical exponent is predicted to be a continuous function of the
quantum control parameter (anisotropy or dimerization) and the
singular behavior of different physical quantities (specific heat,
susceptibility, etc.) are all expected to be related to the value
of the dynamical exponent.

The RG predictions by Fisher\cite{fisherxx} and others\cite{hyman}
have been confronted with results of numerical
studies\cite{haas,stolzexx,girvin}, especially in the RS phase of
isotropic chains, but some cross-over functions of correlations have
also been studied in the Griffiths phase.  In the RS phase some numerical
results are controversial: in earlier studies\cite{stolzexx} a
different scenario from the RG picture is proposed (in particular with
respect to the transverse correlation function), later investigations
on larger finite systems have found satisfactory agreement with the RG
predictions\cite{girvin}, although the finite-size effects were still
very strong.

In the present paper we extend previous work in several directions.
Here we consider open chains and study both bulk and surface
quantities, as well as end-to-end correlations. We develop a
phenomenological theory which is based on the scaling properties of
rare events and determine the complete set of critical decay
exponents. We calculate numerically (off-diagonal) spin-operator
profiles, whose scaling properties are related to (bulk and
surface) decay exponents\cite{turbanigloi} and compare the profiles
with predictions of conformal invariance. Another new feature of our
work is the study of dynamical correlations, both at the critical point
and in the Griffiths phase.  Finally, we perform a detailed analytical
and numerical study of the Griffiths phase and calculate, among
others, the exact value of the dynamical exponent, $z$ in
Eq.(\ref{z}).

The structure of the paper is the following. The model and its
free-fermion representation are presented in Section 2. A
phenomenological theory based on the scaling behavior of rare events
is developed in Section 3. Results in the critical state, where
there is quasi-long-range order in the
chains is presented in Section 4,
whereas the Griffiths phase is studied in Section 5. We discuss the
extensions of our results to random antiferromagnetic $XXZ$ chains and
to higher dimensions in the final Section, whereas some technical
calculations are presented in the Appendices.

\section{The model and its free-fermion representation}

\subsection{The $XY$ and $XX$ models}

We consider an open $XY$ chain (i.e. with free boundary conditions)
with $L$ sites described by the Hamiltonian:
\be
H=\sum_{l=1}^{L-1}\left( J_l^x S_l^x S_{l+1}^x+ J_l^y S_l^y S_{l+1}^y
\right)\;,
\label{hamilton}
\ee
where the $S_l^{\mu}$ ($\mu=x,y$) are spin-$1/2$ operators and the
couplings ($J_l^{\mu}>0$) are independent random variables with distributions
$\pi^{\mu}(J^{\mu})$. The quantum control parameter is the average
anisotropy defined as:
\be
\delta_a={[\ln J^x]_{\rm av} - [\ln J^y]_{\rm av} \over {\rm var}[\ln J^x]+{\rm var}[\ln J^y]}\;,
\label{aniz}
\ee
where ${\rm var}(x)$ is the variance of random variable $x$ and $[\dots]_{\rm av}$
denotes average over quenched
disorder. For $\delta_a>0~(<0)$ there is long-range-order in the $x$ ($y$)
direction, i.e. $\lim_{r \to \infty} [C^{\mu}(r)]_{\rm av} \ne 0$, where
\be
[C^{\mu}(r)]_{\rm av}=[\langle 0 |S_l^{\mu} S_{l+r}^{\mu}|0 \rangle]_{\rm av}\;,
\label{spat}
\ee
and for $\delta_a=0$ the system is in a critical state with
quasi-long-range-order, where correlations decay algebraically, i.e.
\be
[C^{\mu}(r)]_{\rm av} \sim r^{-\eta^{\mu}}\;.
\label{Ccr}
\ee
In the $XX$ model, where the $x$ and $y$ couplings are correlated as
$J_l^x=J_l^y=J_l$, we introduce alternation such that even ($e$) and
odd ($o$) couplings, connecting the site $2i,2i+1$ and $2i-1,2i$,
respectively, are taken from distributions $\rho^e(J_e)$ and
$\rho^o(J_o)$, respectively. For the $XX$ model the quantum control
parameter is the average dimerization defined as:
\be
\delta_d={[\ln J_o]_{\rm av} - [\ln J_e]_{\rm av} \over {\rm var}[\ln J_o]+{\rm var}[\ln J_e]}\;.
\label{dimer}
\ee
The RS phase is at $\delta_d=0$, whereas $\delta_d \ne 0$ corresponds
to the random dimer (RD) phase.  Throughout the paper we use two types
of random distributions, both for the $XY$ and $XX$ models. For the
$XY$ model with the binary distribution the $J^x$ couplings can take
two values $\lambda>1$ and $1/\lambda$ with probability $p$ and $q=1-p$,
respectively, while the couplings $J^y$ are constant:
\beqn
\pi^x(J^x) & = & p \delta(J^x-\lambda)+ q \delta(J^x-\lambda^{-1});\nonumber\\
\pi^y(J^y) & = & \delta(J^y-J^y_0)\;.
\label{binary}
\eeqn
At the critical point $(p-q)\ln \lambda= \ln J^y_0$.  The uniform
distribution is defined via
\beqn
\pi^x(J^x) & = &\cases{1,&for $0<J^x<1$\cr
                0,&otherwise\cr}\;\nonumber\\
\pi^y(J^y) & = &\cases{(J^y_0)^{-1},&for $0<J^y<J^y_0$\cr
                0,&otherwise\cr}\;.
\label{uniform}
\eeqn
and the critical point is at $J^y_0=1$.

For the $XX$ model the corresponding distributions $\rho^e(J^e)$ and
$\rho^o(J^o)$ follows from the correspondences:
\beqn
& J^x \to J^e,~~J^y \to J^o,\nonumber\\
&\pi^x(J^x) \to \rho^e(J^e),~~
\pi^y(J^y) \to \rho^o(J^o)\;.
\label{xyxx}
\eeqn

Note that the critical points of the two models ($\delta_a=0$ and
$\delta_d=0$, respectively) are not equivalent due to the different disorder correlations.

\subsection{The $XY$ chain and the directed walk model}

Using the Jordan-Wigner transformation, the $XY$ model Hamiltonian in
Eq.(\ref{hamilton}) can be rewritten as a quadratic form in fermion
operators. It is then diagonalized through a canonical transformation
which gives
\be
H=\sum_{q=1}^L \epsilon_q(\eta_q^+ \eta_q - {1 \over 2})\;.
\label{fermion}
\ee
The fermion excitations are non-negative and satisfy the set of equations
\beqn
\epsilon_q \Psi_q(l)&=&J^y_{l-1} \Phi_q(l-1) + J_l^x \Phi_q(l+1)\nonumber\\
\epsilon_q \Phi_q(l)&=&J^x_{l-1} \Psi_q(l-1) + J_l^y \Psi_q(l+1);\,
\label{psiphi}
\eeqn
with the boundary conditions $J^x_L=J^y_L=0$. The vectors $\Phi_q$'s
and $\Psi_q$'s which are related to the coefficients of the canonical
transformation are normalized. They enter into the expressions of correlation
functions and thermodynamic quantities.

\begin{figure}
\epsfxsize=\columnwidth\epsfbox{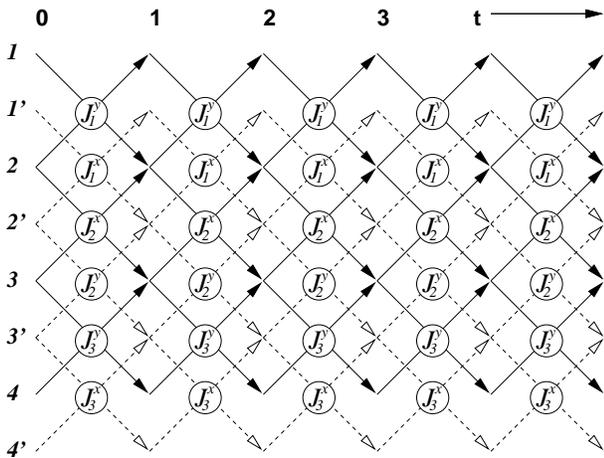}
\caption{Sketch of the directed walk problem corresponding to the
  transfer matrix given in eq. (\protect{\ref{tmatrix}}). Note that
  one has two independent walks in the ${\bf t}$-direction of the
  diagonally layered square lattice, corresponding to the independent
  subspaces for the eigenvalue problem (see text). The coupling
  strength $J_i^{x,y}$ are the transition rates for the random
  walker from one sites in row $i$ to those in row $i\pm1$.}
\label{fig1}
\end{figure}

Usually one proceeds\cite{liebetal} by eliminating either $\Psi_q$ or $\Phi_q$ in
Eqs.(\ref{psiphi}) and the excitations are deduced from the solution
of quadratic equations. This last step can be avoided by introducing
a $2L$-dimensional vector $V_q$ with components:
\beqn
& V_q(4l-3)=\Phi_q(2l-1),~~~& V_q(4l-2)=\Psi_q(2l-1),\nonumber\\
& V_q(4l-1)=\Psi_q(2l),~~~  & V_q(4l)=\Phi_q(2l);\,
\label{Vq}
\eeqn
and noticing that the relations in Eqs.(\ref{psiphi}) then correspond to
the eigenvalue problem of the matrix:
\end{multicols}
\widetext
\noindent\rule{20.5pc}{.1mm}\rule{.1mm}{2mm}\hfill
\be
T = \left(
\matrix{
 0    &    0  & J_1^y &       &           &           &           &           &           &     \cr
 0    &    0  &  0    & J_1^x &           &           &           &           &           &     \cr
J_1^y &    0  &  0    & 0     & J_2^x     &           &           &           &           &     \cr
      & J_1^x &  0    & 0     &  0        & J_2^y     &           &           &           &     \cr
      &       & J_2^x &    0  &  0        &  0        &  J_3^y    &           &           &     \cr
      &       &       & J_2^y &  0        &  0        &    0      &\ddots     &           &     \cr
      &       &       &       &    \ddots &\ddots     &\ddots     & \ddots    & J_{L-1}^y &     \cr
      &       &       &       &           &J_{L-2}^y  &  0        &    0      &   0       &  J_{L-1}^x\cr
      &       &       &       &           &           & J_{L-1}^y &    0      &   0       &   0 \cr
      &       &       &       &           &           &           & J_{L-1}^x &    0      &   0 \cr}
\right)
\label{tmatrix}
\ee
\hfill\rule[-2mm]{.1mm}{2mm}\rule{20.5pc}{.1mm}
\begin{multicols}{2} 
\narrowtext
\noindent 
The matrix $T$ can be interpreted as the transfer matrix (TM) of a
directed walk (DW) problem on four inter-penetrating, diagonally
layered square lattices. Each walker makes steps with weights $J_l^x$
and $J_l^y$ between next-neighbor sites on one of the four square
lattices and the walk is directed in the diagonal direction (see Fig.
\ref{fig1}).

According to Eqs.(\ref{psiphi}), changing $\Phi_q$ into $-\Phi_q$ in
$V_q$, the eigenvector corresponding to $-\epsilon_q$ is obtained.
Thus all information about the DW and the XY model is contained in
that part of the spectrum with $\epsilon_q \ge 0$.  Later on we shall
consider this sector. We note that similar correspondence has been
established earlier between the DW and the transverse-field Ising
model (TIM)\cite{igloiturban96}.

The eigenvalues of $T$ in (\ref{tmatrix}) are of two classes. For
$q=2i-1,~i=1,2,\dots,L$ the odd components of the eigenvectors are
zero, i.e. $V_{2i-1}(2j)=0,~j=1,2,\dots,L$, whereas for the other
class with $q=2i$ the even components are zero, $V_{2i}(2j-1)=0$.
Consequently $T$ can be expressed as a direct product $T=T_{\sigma}
\bigotimes T_{\tau}$, where the trigonal matrices $T_{\sigma}$ ,
$T_{\tau}$ of size $L \times L$ represent transfer matrices of
directed walks. As a result one has to diagonalize these two matrices
of size $L \times L$.  Thus for chains with even number of sites,
$L=2N$, the two classes of eigenvectors are given in terms of the
variables $\Phi$ and $\Psi$ via:
\be
\ba{llclcl}
\epsilon_{2k-1}:\quad  &\Phi_{2k-1}(2j)&=&\Psi_{2k-1}(2j-1)&=&0\\
\epsilon_{2k}:  \quad  &\Phi_{2k}(2j-1)&=&\Psi_{2k}(2j)&=&0
\ea
\label{phizero}
\ee
for $i,j=1,\ldots,N$. Furthermore we assume that the vectors
$\underline{\Phi}_q$and $\underline{\Psi}_q$ are normalized to 1
separately.
 
For the XX model the even and odd sectors are degenerate,
$\epsilon_{2k-1}=\epsilon_{2k}$, thus it is sufficient to diagonalize
only one matrix. In this case one has the additional relations:
\be
\ba{lcl}
\Phi_{2k-1}(2j-1)&=&\Psi_{2k}(2j-1)\;,\\
\Phi_{2k}(2j)    &=&\Psi_{2k-1}(2j)\;.
\ea
\quad{\rm XX-model}\label{XXrel}
\ee
The matrices $T_{\sigma}$ and $T_{\tau}$ are in one-to-one
correspondence with the eigenvalue problem of one-dimensional TIM-s.
This exact mapping for finite open chains is presented in Appendix A.

\subsection{Local order-parameters}

Next we are going to study the long-range-order in the ground state of
the system.  Having free boundary conditions, as in (\ref{hamilton}),
the expectation value of the local spin operator $\langle 0 |
S_l^x|0\rangle$ (and $\langle 0 | S_l^y|0\rangle$) is zero for finite
chains. Then the scaling behavior of the spin operator can be obtained
from the asymptotic behavior of the (imaginary) time-time correlation
function:
\beqn
G_l^x(\tau)&=&\langle 0 | S_l^x(\tau) S_l^x(0) |0\rangle\nonumber\\
&=&\sum_{\langle n|} |\langle n| S_l^x | 0 \rangle |^2 \exp[-\tau
(E_n-E_0)]\;,
\label{Gx}
\eeqn
where $|0\rangle$ and $|n\rangle$ denote the ground state and the
$n$th excited state of $H$ in Eq. (\ref{hamilton}), with energies
$E_0$ and $E_n$, respectively. In the phase with long-range-order the
first excited state is asymptotically
degenerate with the ground state in the thermodynamic limit, thus the sum in Eq. (\ref{Gx}) is
dominated by the first term. In the large $\tau$ limit $\lim_{\tau \to
  \infty} G_l^x(\tau)=(m_l^x)^2$, thus the local order-parameter is
given by the off-diagonal matrix element:
\be
m_l^x=\langle 1 | S_l^x | 0 \rangle\;.
\label{mx}
\ee
In the free fermion representation $S_l^x$ is expressed as\cite{liebetal}
\be
S_l^x={1 \over 2} A_1 B_1 A_2 B_2 \dots A_{l-1} B_{l-1} A_l\;
\ee
with
\be
\ba{lcl}
A_i&=&\sum_{q=1}^L \Phi_q(i)(\eta_q^+ +\eta_q)\;,\\
B_i&=&\sum_{q=1}^L \Psi_q(i)(\eta_q^+ -\eta_q)\;.
\ea
\label{ab}
\ee
Using $|1\rangle = \eta_1^+|0 \rangle$ the matrix-element in Eq.
(\ref{mx}) is evaluated by Wick's theorem. Since for $i \ne j~~\langle
0 | A_i A_j | 0 \rangle= \langle 0 | B_i B_j | 0 \rangle=0$ we
obtain for the local order-parameter:
\be
m_l^x={1 \over 2} \left|\,\matrix{
H_1&G_{11}&G_{12}&\ldots&G_{1l-1}\cr
H_2&G_{21}&G_{22}&\ldots&G_{2l-1}\cr
\vdots&\vdots&\vdots&\ddots&\vdots\cr
H_l&G_{l1}&G_{l2}&\ldots&G_{ll-1}\cr}
\right|\;,
\label{mx1}
\ee
where
\beqn
H_j   &=&\langle0|\eta_1 A_j|0\rangle=\Phi_1(j)\nonumber\\
G_{jk}&=&\langle0|B_k A_j|0\rangle=-\sum_{q} \Psi_q(k) \Phi_q(j)\;.
\label{HG}
\eeqn
For surface spins the local order-parameter is simply given by
$m_1^x=\Phi_1(1)/2$, which can be evaluated in the thermodynamic limit
$L \to \infty$ in the phase with long-range-order, when
$\epsilon_1=0$.  Using the normalization condition $\sum_l
|\Phi_1(l)|^2=1$ we obtain for the surface order parameter:
\be
\ba{lcll}
m_1^x&=&\displaystyle
{1 \over 2} \left[1+\sum_{l=1}^{L/2-1} \prod_{j=1}^l
\left( J^y_{2j-1} \over J^x_{2j} \right)^2\right]^{-1/2}&\quad XY\\
m_1^x&=&\displaystyle
{1 \over 2} \left[1+\sum_{l=1}^{L/2-1} \prod_{j=1}^l
\left( J_{2j-1} \over J_{2j} \right)^2\right]^{-1/2}&\quad XX.
\ea
\label{peschel}
\ee
We note that this formula is {\it exact} for finite chains if we use
fixed spin boundary condition, $S_L^x=\pm 1/2$, wich amounts to have
$J_{L-1}^y=0$. In the fermionic description the two-fold degeneracy of
the energy levels, corresponding to $S_L^x=1/2$ and $S_L^x=-1/2$, is
manifested by a zero energy mode in Eq. (\ref{fermion}) and from the
corresponding eigenvector one obtains $m_1^x$ in Eq. (\ref{peschel})
for any finite chain.

For non-surface spins the expression of the local order-parameter in
Eq. (\ref{mx1}) can be simplified by using the relations in
(\ref{phizero}). Then, half of the elements of the determinant in
(\ref{mx1}) are zero, the non-zero-elements being arranged in a
checker-board pattern, and $m_l^x$ can be expressed as a product of
two determinants of half-size, which reads for $l=2j$ as:
\beqn
m_{2j}^x=&\displaystyle{1 \over 2}& \left|\,\matrix{
H_1&G_{1,2}&G_{1,4}&\ldots&G_{1,2j-2}\cr
H_3&G_{3,2}&G_{3,4}&\ldots&G_{3,2j-2}\cr
\vdots&\vdots&\vdots&\ddots&\vdots\cr
H_{2j-1}&G_{2j-1,2}&G_{2j-1,4}&\ldots&G_{2j-1,2j-2}\cr}
\right|\nonumber\\
&\times&
\left|\,\matrix{
G_{2,1}&G_{2,3}&\ldots&G_{2,2j-1}\cr
G_{4,1}&G_{4,3}&\ldots&G_{4,2j-1}\cr
\vdots&\vdots&\ddots&\vdots\cr
G_{2j,1}&G_{2j,3}&\ldots&G_{2j,2j-1}\cr}
\right|
\;.
\label{mx2}
\eeqn
The local order-parameter $m_l^y$, related to the off-diagonal
matrix-element of the operator $S_l^y$ can be obtained from Eqs.
(\ref{mx1}) and (\ref{peschel}) by exchanging $J_l^x \leftrightarrow
J_l^y$.

For the $S^z_l$ operator the autocorrelation function, $G^z_l(\tau)$,
can be expressed in a similar way as $G^x_l(\tau)$ in Eq. (\ref{Gx})
and its long time limit, $\lim_{\tau \to \infty}
G_l^z(\tau)=(m_l^z)^2$, is given by the local order parameter
\be
m_l^z=\langle \phi_z | S_l^z | 0 \rangle\;.
\label{mlz}
\ee
Here $| \phi_z \rangle$ denotes the lowest eigenstate of $H$ in Eq.
(\ref{hamilton}) having a non-vanishing matrix-element of $S_l^z$ with
the ground state.  In the free fermion representation $S_l^z$ can be
written as\cite{liebetal}
\be
S_l^z={1 \over 2} A_l B_l\;
\label{slz}
\ee
and the off-diagonal order-parameter is given by:
\be
m_l^z= {1 \over 2} |-\Phi_1(l) \Psi_2(l) + \Psi_1(l) \Phi_2(l)|\;.
\label{mz}
\ee
For the $XX$ model one can obtain simple expressions using the
relations in Eqs. (\ref{XXrel}) as:
\beqn
m_{2i-1}^z&=\frac{1}{2}[\Phi_1(2i-1)]^2\;,&\nonumber\\
&&\quad XX{\rm-model}\label{mzxx}\\
m_{2i}^z  &=\frac{1}{2}[\Psi_1(2i)]^2\;.\nonumber
\eeqn

\subsection{Autocorrelations}

Next we consider the dynamical correlations of the system as a
function of the imaginary time $\tau$. First, we note that the
correlations between $x$-components of the surface spins can be
obtained directly from eq(\ref{Gx}) as:
\beqn
G_1^x(\tau)&=&{1 \over 4} \sum_q | \Phi_q(1)|^2 \exp(-\tau \epsilon_q)
\nonumber\\
&=&{1 \over 4} \sum_i^{L/2} | \Phi_{2i-1}(1)|^2 \exp(-\tau \epsilon_{2i-1})\;,
\label{surfcorr}
\eeqn
where we have used the relations in Eq. (\ref{phizero}).

For bulk spins the matrix-element $\langle n|S_l^x|0\rangle$ in
eq(\ref{Gx}) is more complicated to evaluate, one has to go
back to the first equation of (\ref{Gx}) and considers the
time-evolution in the Heisenberg picture:
\beqn
S_l^x(\tau)&=&\exp(\tau H) S_l^x \exp(-\tau H)\nonumber\\
&=&{1 \over 2}A_1(\tau) B_1(\tau)
\dots A_{l-1}(\tau) B_{l-1}(\tau) A_l(\tau)\;.
\label{sigmatau}
\eeqn

The general time and position dependent correlation function
\be
\langle S_l^x(\tau) S_{l+n}^x\rangle
={1 \over 4}\langle A_1(\tau) B_1(\tau) \cdots A_l(\tau) A_1 B_1\dots A_{l+n}\rangle\;,
\label{gencorr}
\ee
can then be expanded using Wick's theorem into a sum over products of
two-operator expectation values, which can be expressed in a
compact form as a Pfaffian:
\end{multicols}
\widetext
\noindent\rule{20.5pc}{.1mm}\rule{.1mm}{2mm}\hfill
\beqn
4\langle S_l^x(\tau) S_{l+n}^x\rangle
&=&\left.
\matrix{|\;
\langle A_1(\tau)B_1(\tau)\rangle &
\langle A_1(\tau)A_2(\tau)\rangle &
\langle A_1(\tau)B_2(\tau)\rangle &
\cdots\quad
\langle A_1(\tau)A_l(\tau)\rangle &
\langle A_1(\tau)A_1\rangle &
\cdots\quad
\langle A_1(\tau)A_{l+n}\rangle\cr
&
\langle B_1(\tau)A_2(\tau)\rangle &
\langle B_1(\tau)B_2(\tau)\rangle &
\cdots\quad
\langle B_1(\tau)A_l(\tau)\rangle &
\langle B_1(\tau)A_1\rangle &
\cdots\quad
\langle B_1(\tau)A_{l+n}\rangle\cr
& &
\langle A_2(\tau)B_2(\tau)\rangle &
\cdots\quad
\langle A_2(\tau)A_l(\tau)\rangle &
\langle A_2(\tau)A_1\rangle &
\cdots\quad
\langle A_2(\tau)A_{l+n}\rangle\cr
& &         & \ddots & & \vdots\cr
& &         & & &  \langle B_{l+n-1}A_{l+n}\rangle}
\right\vert\nonumber\\
&=&
\pm \left[ {\rm det}\, C_{ij}\right]^{1/2}\;,
\label{pfaffian}
\eeqn
\begin{multicols}{2} 
\narrowtext
\noindent 
where $C_{ij}$ is an antisymmetric matrix $C_{ij}=-C_{ji}$, with the
elements of the Pfaffian (\ref{pfaffian}) above the diagonal.
%
%
%
%
At zero temperature the elements of the
Pfaffian are the following:
\begin{eqnarray}
\langle A_j(\tau) A_k\rangle
& = &\sum_q \Phi_q(j) \Phi_q(k) \exp(-\tau \epsilon_q)\;,
\nonumber\\
\langle A_j(\tau) B_k\rangle
& = &\sum_q \Phi_q(j) \Psi_q(k) \exp(-\tau \epsilon_q)\;,
\nonumber\\
\langle B_j(\tau) B_k\rangle
& = &-\sum_q \Psi_q(j) \Psi_q(k) \exp(-\tau \epsilon_q)\;,
\nonumber\\
\langle B_j(\tau) A_k\rangle
& = &-\sum_q \Psi_q(j) \Phi_q(k) \exp(-\tau \epsilon_q)\;,
\label{pfaffelm}
\end{eqnarray}
whereas the equal-time contractions are given below (\ref{ab}). For the
finite temperature contractions see c.f. \cite{stolze}.

For longitudinal correlations the matrix-elements of $S_l^z$ in
(\ref{slz}) is given in a simple form for any position, $l$, therefore
$G_l^z(\tau)$ can be obtained from the analogous expression to
(\ref{Gx}) as
\beqn
G_l^z(\tau)={1 \over 4} \sum_{q} \sum_{p > q}
&\;&|-\Psi_{p}(l)\Phi_{q}(l) +\Psi_{q}(l)\Phi_{p}(l)|^2
\nonumber\\
&\cdot&\exp[-\tau(\epsilon_{q}+\epsilon_{p})]\;.
\label{Glz}
\eeqn

\section{Phenomenological theory from scaling of rare events}

In classical random ferromagnets where the critical behavior is
controlled by a random fixed point the distribution of several
physical quantities (order-parameters, correlations,
autocorrelations, etc.) is broad and as a consequence these quantities
are not self-averaging: their average and most-probable or typical
values are different. In random quantum spin chains the critical
properties are expected to be controlled by the infinite-randomness
critical fixed point\cite{fisherb}, where the distributions are
extremely (logarithmically) broad and as a consequence the average and
typical behavior of these quantities are completely different.  The
average is dominated by such realizations (the so called {\it rare
  events}), which have a very large contribution, but their fraction
is vanishing in the thermodynamic limit. In this Section we identify
these rare events for the random $XY$ (and $XX$) model and use their
properties to develop a phenomenological theory.  Our basic
observations are related to exact relations about the surface
order-parameter and the energy of low-lying excitations.

\subsection{Surface order-parameter and the mapping to adsorbing random walks}

The local order-parameter at the boundary is given by the simple
formula in Eq. (\ref{peschel}) as a sum of products of the ratio of
the couplings $J^y_{2j-1}$ and $J^x_{2j}$. It is easy to see from Eq.
(\ref{peschel}) that in the thermodynamic limit the average surface
order-parameter is zero (non-zero), if the geometrical mean of the
$J^x_{2j}$ couplings is (smaller) grater than that of the $J^y_{2j-1}$
couplings. From this the definition of the control parameters in Eqs.
(\ref{aniz}) and (\ref{dimer}) follows.

Next we compute the average value of the surface order-parameter for
the extreme binary distribution, i.e.\ the limit $\lambda \to 0$
\cite{extreme} in (\ref{binary}). For a random realization of the
couplings the surface order-parameter at the critical point
($p=q=1/2$) is zero, whenever a product of the form of $\prod_{i=1}^l
(J_i^x)^{-2}$, $l=1,2,\dots,L$ is infinite, i.e. the number of
$\lambda$-couplings exceeds the number of $\lambda^{-1}$-couplings in
any of the $[1,l]$ intervals. Otherwise the surface order-parameter
has a finite value of $O(1)$. The distribution of the couplings $J^x$
can be represented by one-dimensional random walks that start at zero
and make the $i$-th step upwards (for $J^x_{2i}=\lambda^{-1}$) or
downwards (for $J^x_{2i}=\lambda$). The ratio of walks representing a
sample with finite surface order-parameter is given by the survival
probability of the walk $P_{surv}$, i.e.\ the probability of the
walker to stay always above the starting point in $L/2$ steps which is
given by $P_{surv}(L/2) \sim L^{-1/2}$.

Next we consider the vicinity of the critical point, when the scaling
behavior of the average surface order-parameter can be obtained from
the survival probabilities of biased random walks\cite{bigpaper}, where
the probability that the walker makes a step towards the adsorbing
boundary, $q$, is different from that of a step off the boundary, $p$.
The control parameter of the walk, $\delta_w=p-q$, is analogous to the
quantum control parameters $\delta_a$ and $\delta_d$ in Eqs.
(\ref{aniz}) and (\ref{dimer}), respectively.  Thus we have the basic
correspondences between the average surface order-parameter of the
$XY$ (and $XX$) model and the surviving probability of adsorbing
random walks:
\be
[m_1(\delta,L)]_{av} \sim P_{\rm surv}(\delta_w,L/2),~~~\delta 
\sim \delta_w\;,
\label{m1psurv}
\ee
We recall the asymptotic properties of the surviving probability of
adsorbing random walks\cite{bigpaper}. For unbiased walks:
\be
P_{\rm surv}(\delta_w=0,L) \sim L^{-1/2}\;,
\label{ps0}
\ee
for walks with a drift away from the wall:
\be
P_{\rm surv}(\delta_w>0,L \to \infty) \sim \delta_w\;,
\label{ps+}
\ee
and for walks with a drift towards the wall:
\be
P_{\rm surv}(\delta_w<0,L ) \sim \exp(-L/\xi_w),~~~\xi_w \sim \delta_w^{-2}\;.
\label{ps-}
\ee
In this way we have identified the rare events for the surface
order-parameter, which are samples with a coupling distribution which
have a surviving walk character. The scaling properties of the average
surface order-parameter and the correlation length immediately follow
from Eqs. (\ref{ps0}), (\ref{ps+}) and (\ref{ps-}) and will be
evaluated in Section IV.A.

\subsection{Scaling of low-energy excitations}

The rare events controlling the surface order-parameter are also
important for the low-energy excitations. Our results are obtained by
using a simple relation for the smallest gap, $\epsilon_1(l)$, of an
open system of size $l$, i.e.  with free boundary conditions,
expecting that it goes to zero at least as $\sim 1/l$. With this
condition one can neglect the r.h.s. of the eigenvalue problem of
${\bf T}$ in Eq.  (\ref{tmatrix}), ${\bf T} V_1=\epsilon_1 V_1$, and
derive approximate expressions for the eigenfunctions $\Phi_1$ and
$\Psi_1$. With these one arrives at:
\be
\epsilon_1(l) \sim m_1^x m_{l-1}^x J^y_{l-1}\prod_{j=1}^{l/2-1} 
{J^y_{2j-1} \over J^x_{2j}} \;,
\label{epsilon1}
\ee
Here $m_1^x$ is defined in (\ref{peschel}) and the surface
order-parameter at the other end of the chain, $m_{l-1}^x$, is given
as in Eq. (\ref{peschel}) replacing $J^y_{2j-1}/J^x_{2j}$ by
$J^y_{l+1-2j}/J^x_{l-2j}$.  (For details of the derivation of a
similar expression for the quantum Ising chain see in
Ref.\onlinecite{itks}.)

Before using the relation in Eq. (\ref{epsilon1}) we note that
(surface) order and the presence of low-energy excitations are
inherently related. These samples with an exponentially (in the
system size) small gap have finite, $O(1)$, order-parameters at both
boundaries and the coupling distribution follows a surviving walk
picture. Such type of coupling configuration represents a strongly
coupled domain (SCD), which at the critical point extends over the
size of the system, $L$. In the off-critical situation, in the
Griffiths phase the SCD-s have a smaller extent, $l \ll L$, and they
are localized both in the volume and near the surface of the
system. The characteristic excitation energy of an SCD can be
estimated from Eq. (\ref{epsilon1}) as
\be
\epsilon_1(l) \sim \prod_{j=1}^{l/2-1} {J^y_{2j-1} \over J^x_{2j}}
\sim \exp\Bigl\{-\frac{l_{\rm tr}}{2} \overline{\ln(J^y/J^x)}\Bigr\}\;,
\label{epsilon11}
\ee
where $l_{\rm tr}$ measures the size of transverse fluctuations of
a surviving walk of length $l$ and $\overline{\ln(J^x/J^y)}$ is an
average ratio of the couplings, (it is $\overline{\ln(J^e/J^o)}$ for
the $XX$ model).

At the critical point ($\delta=0$), where $l \sim L$, the size
of transverse fluctuations of the couplings in the SCD is $l_{\rm tr}
\sim L^{1/2}$\cite{bigpaper}. Consequently we obtain from Eq. (\ref{epsilon11})
for the scaling relation of the gap:
\be
\epsilon(\delta=0,L) \sim \exp(-{\rm const}\cdot L^{1/2})\;.
\label{scener}
\ee
Then the appropriate scaling variable is $\ln \epsilon/ \sqrt{L}$ and
the distribution of the excitation energy is extremely (logarithmically)
broad.

In the Griffiths phase the size of a SCD can be estimated along the
lines of Ref.\onlinecite{bigpaper} as $l \sim \xi_w \ln L$ and the size
of transverse fluctuations is now $l_{\rm tr} \sim l \sim \ln L$.
Setting this estimate into Eq. (\ref{epsilon11}) we obtain for the
scaling relation of the gap:
\be
\epsilon(L) \sim L^{-z}\;,
\label{zL}
\ee
where $z$ is the dynamical exponent as defined in Eq. (\ref{z}). The
distribution of low-energy excitations can be obtained from the
observation that an SCD can be localized at any site of the chain,
thus $P_L(l) \sim P_L(\ln \epsilon) \sim L$. For a given large $L$
the scaling combination from Eq. (\ref{zL}) is $L\epsilon^{1/z}$, thus
we have:
\be
P(\epsilon) \sim \epsilon^{-1+1/z}\;.
\label{edistr} 
\ee
As already mentioned $z$ is a continuous function of the quantum
control parameter $\delta$ and we are going to calculate its exact
value in Section V.

\subsection{Scaling theory of correlations}

The scaling behavior of critical average correlations is also
inherently connected to the properties of rare events. Here the
quantity of interest is the probability $P^{\mu}(l)$, which measures
the fraction of rare events of the local order-parameter $m_l^{\mu}$.
For the surface order-parameter $m_1^x$ it is given by the surviving
probability, $P^x(1)=P_{surv}$, according to Eq. (\ref{m1psurv}).  We
start with the equal-time correlations in Eq. (\ref{spat}). In a given
sample there should be local order at both reference points of the
correlation function in order to have $C^{\mu}(r)=O(1)$. This is
equivalent of having two SCD-s in the sample which occur with a
probability of $P_2^{\mu}(l,l+r)$, which factorizes for large
separation $\lim_{r \to \infty} P_2^{\mu}(l,l+r)=
P^{\mu}(l)P^{\mu}(L+r)$, since the disorder is uncorrelated. The
probability of the occurrence of a SCD at position $l$, $P^{\mu}(l)$,
has the same scaling behavior as the local order-parameter
$[m^{\mu}_l]_{\rm av}=[\langle \phi_{\mu}|S^{\mu}_l|0\rangle]_{\rm
  av}$, which behaves at a bulk point, $0<l/L<1$, as:
\be
[m^{\mu}_l(L)]_{\rm av} \sim L^{-x^{\mu}}\;,
\label{fss}
\ee
whereas for a boundary point, $l=1$, this relation involves the
surface scaling dimension $x^{\mu}_1$. Consequently $P^{\mu}(l)$
transforms as $P^{\mu}(l/b)=b^{-x^{\mu}}P^{\mu}(l)$ under a scaling
transformation, when lengths are rescaled by a factor $b>1$. Recalling
that for spatial correlations there should be two independent SCD-s we
obtain the transformation law:
\be
[C^{\mu}(r)]_{\rm av}=b^{-2x^{\mu}}[C^{\mu}(r/b)]_{\rm av}\;.
\label{Cscal}
\ee
Now taking $b=r$ one recovers the power-law decay in Eq. (\ref{Ccr})
with the exponent
\be
\eta^{\mu}=2 x^{\mu}\;.
\label{xeta}
\ee
For critical time-dependent correlations the scaling behavior is
different from that in Eq. (\ref{Cscal}). This is due to the fact that
disorder in the time direction is perfectly correlated and the
autocorrelation function in a given sample is $G^{\mu}_l(\tau)=O(1)$,
if there is {\it one} SCD localised at position $l$. Therefore the
average autocorrelation function $[G^{\mu}_l(\ln \tau)]_{\rm av}$
scales as the probability of rare events $P^{\mu}(l)$:
\be
[G^{\mu}_l(\ln \tau)]_{\rm av}=b^{-x^{\mu}} [G^{\mu}_l(\ln \tau/ b^{1/2})]_{\rm av}\;,
\label{Gscal}
\ee
where we have used the relation in Eq. (\ref{e1.1}) between relevant
length and time at the critical point. Taking the length scale as
$b=(\ln \tau)^2$ we obtain for points $l$ in the volume:
\be
[G^{\mu}_l(\tau)]_{\rm av} \sim (\ln \tau)^{-\eta^{\mu}}\;,
\label{gx1}
\ee
whereas for surface spins, $l=1$, one should use the corresponding
surface decay exponent $\eta^{\mu}_1$.

Next we turn to study the scaling properties of the average
correlation functions in the Griffiths phase, i.e. outside the
critical point.  For equal-time correlations in a sample
$C^{\mu}(r)=O(1)$, if the SCD extends over a large distance of $r$,
which according to Eq. (\ref{ps-}) is exponentially improbable.  Thus
the average spatial correlations decay as
\be
[C^{\mu}(r)]_{\rm av} \sim \exp(-r/\xi),~~~\xi \sim \xi_w\;,
\label{corr-}
\ee
where $\xi_w$ is defined in Eq. (\ref{ps-}).  On the other hand the
autocorrelation function in a sample is $G^{\mu}(\tau)=O(1)$, if there
is one SCD localized at $l$, which occurs with a probability of
$P^{\mu}(l) \sim 1/L$.  Consequently the average autocorrelation
function, which scales as $P^{\mu}(l)$, transforms under a scaling
transformation as:
\be
[G^{\mu}_l(\tau)]_{\rm av}=b^{-1}
[G^{\mu}_{l/b}(\tau/b^z)]_{\rm av}~~~\delta>0\;,
\label{grscal}
\ee
where we used the scaling combination $\tau/b^z$ in accordance with
Eq. (\ref{z}).  Now taking $b=\tau^{1/z}$ we obtain
\be
[G^{\mu}_l(\tau)]_{\rm av} \sim \tau^{-1/z}\;,
\label{grdecay}
\ee
for any type of autocorrelations, both in the volume and at the surface.

\section{Critical properties}

Here we consider in detail the random $XY$ and $XX$ chains in the
vicinity of the critical points, as defined in Eqs. (\ref{aniz}) and
(\ref{dimer}), respectively. The off-critical properties of the systems
in the Griffiths phase are presented afterwards in the following
Section.

\subsection{Length- and time-scales}

As we argued in the previous Chapter the average behavior of random
quantum spin chains are inherently related to the properties of the
rare events, which are SCD-s, having a coupling distribution of
surviving RW character. The typical size of an SCD, as given by
$\xi_w$ in Eq. (\ref{ps-}), is related to the average correlation
length of the system, $[\xi]_{\rm av}$.  Then using the
correspondences in Eqs. (\ref{m1psurv}), (\ref{ps-}) and
(\ref{corr-}) we get the relation:
\be
[\xi]_{\rm av} \sim |\delta|^{-\nu},~~~\nu=2\;.
\label{nu2}
\ee
The {\it typical} correlation length, $\xi_{\rm typ}$, as measured by
the average of the logarithm of the correlation function is different
from the {\it average} correlation length. One can estimate the
typical value by studying the formula in Eq. (\ref{peschel}) for the
surface order-parameter, where the products are typically of
$\prod_j(J^y_{2j-1}/J^x_{2j})^2 \sim \exp({\rm const}\cdot|\delta| L)$,
thus $[m_s(L,\delta\langle0)]_{\rm typ} \sim \exp(- {\rm const}\cdot|\delta| L)
\sim \exp(-L/\xi_{\rm typ})$. Thus we obtain:
\be
\nu_{\rm typ}=1\;.
\label{nutyp}
\ee
We note that {\it at the critical point} the largest value of the
above products is typically of $\prod_j(J^y_{2j-1}/J^x_{2j})^2 \sim
\exp( A L^{1/2})$, since the transverse fluctuations in the couplings
are of $O(L^{1/2})$, thus we have $[m_s(L,\delta=0)]_{\rm typ} \sim
\exp(- {\rm const}\cdot L^{1/2})$.

As shown in Eq.  (\ref{epsilon11}) the value of the smallest gap is
related to the size of transverse fluctuations of an SCD, $l_{\rm
  tr}$. Away from the critical point, when the correlation length is
finite, one has $l_{\rm tr} \sim \xi^{1/2}$, and therefore the typical
relaxation time of a sample with typical correlation length $\xi$
scales as
\be
\ln t_r \sim \xi^{1/2}\;.
\label{e1.1}
\ee

We note that the results in this part about length- and time-scales
are valid both for the $XY$ and $XX$ models. They also hold in
identical form for the random TIM\cite{fisher,bigpaper}, which can be
understood as a consequence of the mapping of the $XY$-chain into
decoupled TIM-s.  (See Appendix.) Since the corresponding scaling
expressions for the random TIM have been studied in detail in previous
numerical work\cite{youngrieger,bigpaper} we do not repeat these
calculations here.

\subsection{Quasi-long-range-order}

At the critical point of random quantum chains the equal-time
correlations decay with a power-law, (see Eq. (\ref{Ccr})), thus there is
QLRO in the system. The decay exponent of critical correlations are
related to the scaling exponent $x^{\mu}$ of the fraction of rare
events of the given quantity (see Eq. (\ref{xeta})) and its value
generally depends on the type of correlations of the disorder, thus it
could be different for the $XY$ and the $XX$ models. Analyzing the
scaling properties of the rare events in the $XY$ and $XX$ chains we
have calculated the critical decay exponents of different correlation
functions, both between two spins in the volume and for end-to-end
correlations. Our results are presented in Table I.

\bc
\begin{tabular}{|c||c|c|c|c|}
\hline
&$\eta^x(XY)$&$\eta^x(XX)$&$\eta^z(XY)$&$\eta^z(XX)$\\
\hline
bulk&$3-\sqrt{5}^{(**)}$&$2^{(*)}$&$4$&$2^{(*)}$\\
surface&$1$&$1$&$2$&$1$\\
\hline
\end{tabular}
\ec
{\small
TABLE I: Decay exponents of critical correlations in the random $XY$ 
and $XX$ chains. The exponents with a superscript $^{(*)}$ are those 
calculated by Fisher with the RG
method\cite{fisherxx}, whereas $^{(**)}$ follows from the results of
the random TIM in Ref.\cite{fisher}.}
\medskip

In the following we are going to derive these exponents by analytical
and scaling methods and then confront them with the results of
numerical calculations.

\subsubsection{Longitudinal order-parameter}

We start with the scaling behavior of the longitudinal
order-parameter $m_l^z$, which in the $XX$ chain is given by the simple
formula in Eq. (\ref{mzxx}).  Summing over all sites one gets
the sum-rule
\be
\sum_{l=1}^L m_l^z=1~~~{\rm XX-model}\;,
\label{sumrule}
\ee
where we have used Eq. (\ref{phizero}) and the fact that the $\Phi_q$
and $\Psi_q$ are normalized.  Since this sum-rule is valid for the
average quantities, too, we get immediately
\be
[m^z_l]_{\rm av}=L^{-1}\tilde{m}^z(l/L)\;,
\ee
where $\tilde{m}^z(\tilde{l})$ is a scaling function with $\tilde{l}=l/L$.
Consequently for bulk spins the finite-size dependence of the local
order-parameter is $[m^z_l]_{\rm av} \sim L^{-1}$, thus from Eq.
(\ref{fss}) we have $x^z(XX)=1$ and from Eq. (\ref{xeta}) the decay
exponent is 
$$\eta^z(XX)=2$$
as given in Table I. A further consequence
of the sum-rule in Eq. (\ref{sumrule}) is that the average value of
the bulk order-parameter is the same, if the averaging is performed
over any single sample. Thus the order-parameter $m^z$ and the
correlation function $\langle S^z_l S^z_{l+r} \rangle$ are {\it
self-averaging}. This is quite special in disordered systems where the
correlations are generally not self-averaging\cite{derrida}. The
self-averaging properties of the $S^z$-correlations provides an
explanation of the accurate numerical determination of the decay
exponent $\eta^z(XX)=2$ in previous numerical
work\cite{stolzexx,girvin}.

The surface order-parameter $m_1^z$ for the $XX$ model satisfies the
relation $m_1^z = 2 \left( m_1^x\right)^2$, which follows from Eqs.
(\ref{peschel}) and (\ref{mzxx}). Then a rare event with $m_1^x=O(1)$
is also a rare event for the order-parameter $m_1^z$, consequently the
fraction of rare-events $P^z_1$ is given by the surviving probability
in Eq. (\ref{ps0}). Thus the scaling dimension is $x^z_1=1/2$ and the
decay exponent of critical end-to-end correlations is 
$$\eta^z_1(XX)=1$$ 
as shown in Table I.

\begin{figure}
\epsfxsize=\columnwidth\epsfbox{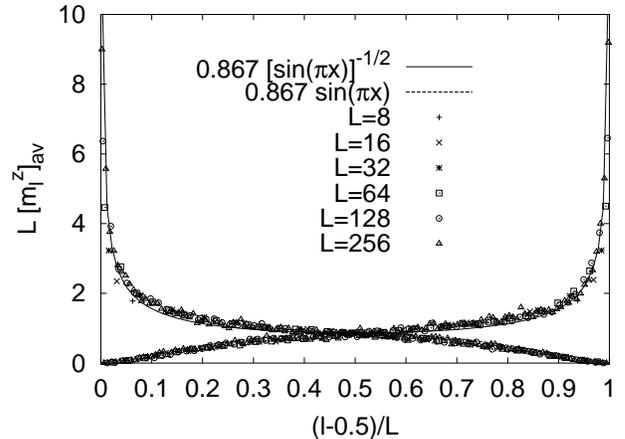}
\caption{Finite size scaling plot of the longitudinal
  order-parameter profiles $[m_l^z]_{\rm av}$ for the $XX$ model at
  criticality for different system sizes calculated numerically with
  the fermion method using eq.  (\protect{\ref{mzxx}}). The data are
  for the uniform distribution, averaged over 50000 samples. The conformal
  results are indicated by full lines.}
\label{fig2}
\end{figure}

We studied the order-parameter profile $[m_l^z]_{\rm av}$
numerically for large finite systems up to $L=256$. As shown in Fig.
\ref{fig2} the numerical points of the scaled variable $L[m^z_l]_{\rm
  av}$ are on one scaling curve $\tilde{m}^z(\tilde{l})$ for different
values of $L$.  The scaling curve has two symmetric branches for odd
and even lattice sites, which cross at $l=L/2$.  The upper part of the
curves in the large $L$ limit is very well described by the function
$\tilde{m}^z(\tilde{l})_{\rm u}={\cal A} \sin(\pi \tilde{l})^{-1/2}$,
which corresponds to the conformal result about off-diagonal
matrix-element profiles\cite{turbanigloi}:
\be
[m_l^{\mu}]_{\rm av} \sim \left({\pi \over L}\right)^{x^{\mu}} 
\left( \sin \pi {l \over L}\right)^{x^{\mu}_1-x^{\mu}}\;,
\label{conf}
\ee
with $x^z=1$ and $x^z_1=1/2$. On the other hand the lower part of the
curves in Fig.\ \ref{fig2} is given by $\tilde{m}^z(\tilde{l})_{\rm
  l}={\cal A} \sin(\pi \tilde{l})$, which corresponds to Eq.
(\ref{conf}) with $x^z_2=2$. Thus we obtain that average critical
correlations between two spins which are next to the surface are
decaying as $[C^z(2,L-1)]_{\rm av} \sim L^{-4}$.  Using the sum-rule
for the profile in Eq. (\ref{sumrule}) and the conformal predictions
one can determine the pre-factor $\cal{A}$ from normalization. Then
from the equation ${\cal A}/2 \int_0^1 [(\sin \pi x)^{-1/2}+ \sin \pi
x] d x=1$, one gets ${\cal A}=.86735$, which fits well the numerical
data on Fig. \ref{fig2}.

These results about the conformal properties of the profile are in
agreement with similar studies of the random
TIM\cite{profiles,bigpaper}. Thus it seems to be a general trend that
critical order-parameter profiles of random quantum spin chains are
described by the results of conformal invariance, although these
systems are strongly anisotropic (see Eq. (\ref{e1.1})) and therefore
not conformally invariant.

Next we turn to study the order-parameter $m_l^z$ and the longitudinal
correlation function in the random $XY$ model. In this model the
disorder in the $J^x_l$ and $J^y_l$ couplings is uncorrelated,
therefore one can perform averaging in the two subspaces $T_{\sigma}$
and $T_{\tau}$, or in the two decoupled TIM-s, independently. Note
that the expression for $m_l^z$ in Eq. (\ref{mz}) is given as a
product of two vector-components, where each vector belongs to
different subspaces and have the same average behavior. Since the
couplings entering the two separate eigenvalue problems are
independent one gets for the disorder average
\be
[m_l^z]_{\rm av}=[\Phi_1(l)]_{\rm av}\cdot[\Psi_2(l)]_{\rm av}\;.
\ee
Since the probability for $m_l^z$ being of order one is the product of
the probabilities for $\Phi_1(l)$ and $\Psi_2(l)$ being of order one
we conclude that the scaling dimension for $m_l^z$ in the random $XY$
chain is twice that for the random $XX$ chain. Thus the decay
exponents are 
$$\eta^z(XY)=4$$ 
and 
$$\eta^z_1(XY)=2$$ 
in the bulk and at the surface, respectively, as shown in Table I.

\begin{figure}
\epsfxsize=\columnwidth\epsfbox{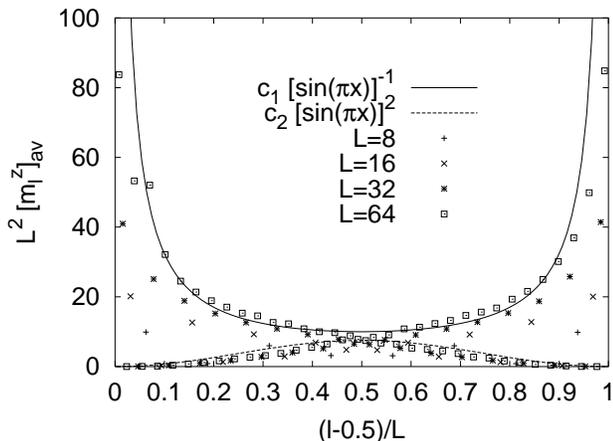}
\caption{Finite size scaling plot of the longitudinal
  order-parameter profiles $[m_l^z]_{\rm av}$ for the $XY$ model at
  criticality for different system sizes calculated numerically with
  the fermion method using eq.  (\protect{\ref{mzxx}}). The data are
  for the uniform distribution, averaged over 50000 samples.}
\label{fig3}
\end{figure}

The numerical results about the order parameter profile is shown in
Fig.\ \ref{fig3}. The data collapse is satisfactory, although not as
good as for the $XX$ model. Similar conclusion holds for the relation
with the conformally predicted profile, which is also presented in
Fig.\ \ref{fig3}.

\subsubsection{Transverse order-parameter}

We start with the surface order-parameter, $m_1^x$, as given by
the simple formula in Eq. (\ref{peschel}). This formula is identical
both for the $XY
$ and $XX$ models and its average behavior follows
from the adsorbing random walk mapping in Section III.A.  Then from
Eqs. (\ref{m1psurv}) and (\ref{ps0}) one gets $x^x_1=1/2$ and
$$\eta^x_1=1\;,$$
both for the random $XY$ and $XX$ models, as shown in
Table I. The value of the decay exponents follows also from the
mapping to two TIM-s. As shown in Eq. (\ref{corrtr1}) in the Appendix
the correlation function $\langle S_{2l}^x S_{2l+2r}^x \rangle$ is
expressed as the product of spin correlations in the two TIM's, one
with open boundary conditions, but the other is taken with fixed-spin
boundary conditions in terms of dual variables. For end-to-end
correlations this second factor in the product is unity, since it is
the correlation between two fixed spins. Therefore end-to-end
correlations between the random TIM and the random $XY$ and $XX$
models are identical and the decay exponent corresponds to the value
in Table I.

\begin{figure}
\epsfxsize=\columnwidth\epsfbox{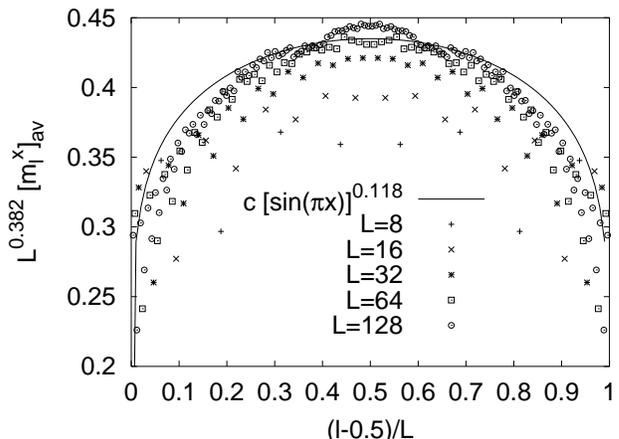}
\caption{Transverse order-parameter profile $[m_l^x]_{\rm av}$ for the
  $XY$ model at criticality for different system sizes calculated
  numerically with the fermion method using eq.
  (\protect{\ref{mx2}}). The data are for the uniform distribution,
  averaged over 50000 samples.  }
\label{fig4}
\end{figure}

For bulk correlations one can easily find the answer for the $XY$
model with the mapping in Eq. (\ref{corrtr1}). When the two points of
reference are located far from the boundary the boundary condition
does not matter and after performing the independent averaging for the
two factors of the product one obtains $[\langle S_{2l}^x S_{2l+2r}^x
\rangle]_{\rm av} =1/4 [\langle \sigma^x_l \sigma^x_{l+r}\rangle]_{\rm
  av}^2$, thus 
\be
\eta^x(XY)=2 \eta(TIM)=3-\sqrt{5}\;,
\label{gold}
\ee
where the last result follows from Fisher's RG
calculation\cite{fisher}. (As shown in Ref.\onlinecite{avpers} the
rare events for the bulk order-parameter in the TIM are samples having
a coupling distribution of average persistence character).
The scaling exponent $x^x(XY)$ can identically be obtained from the
expression of the order-parameter profile in Eq.  (\ref{mx2}), which
is in the form of a product of the two Ising order-parameters and for
the $XY$ model the two factors are averaged independently.

For the $XY$ model the numerically calculated profile is shown
in Fig.\ \ref{fig4}. The scaling plot with the exponents in Table I
is reasonable, although larger systems and even more samples would be
needed to reach the expected asymptotic behavior, as predicted by
conformal invariance in Eq. (\ref{conf}).

The arguments leading to the prediction (\ref{gold}) for the
transverse bulk order parameter exponent do not apply for the $XX$
model and one cannot obtain a simple estimate for the bulk decay
exponent from Eqs. (\ref{corrtr1}) or (\ref{mx2}) due to the
following reason. The expressions with the parameters of the two
quantum Ising chains contain real and dual variables for the two
($\sigma$ and $\tau$) systems. Since $J^x_l=J^y_l=J_l$ a domain of
strong couplings in the $\sigma$ chain corresponds to a domain of weak
couplings in the $\tau$ chain and reverse. Therefore the rare events
of the TIM can not be simply related to the rare-events of the $XX$
chain.

The value for $\eta^x(XX)$, however, can be obtained by the following
argument. For simplicity let us consider the extreme binary
distribution in which $J_{2i}=1$ and $J_{2i-1}=\lambda$ or $1/\lambda$
with probability $1/2$, taking the limit $\lambda\to0$. Then, from
Eq.(\ref{peschel}), one gets only then a non-vanishing transversal
surface magnetization, when the disorder configuration has a surviving
walk character (meaning $\prod_{i=1}^lJ_{2i-1}<\infty$ for all
$l=1,\ldots,L/2-1$). This implies, also for general distributions of
couplings that $m_1^x\sim{\cal O}(1)$ only if the surface spin is {\it
  weakly} coupled to the rest of the system.  It is instructive to
note the difference to the surface magnetization in the TIM, where
$m_1^x\sim{\cal O}(1)$ when the surface spin is {\it strongly} coupled
to the rest of the system, meaning that
$\prod_{i=1}^l(1/J_{i})<\infty$ for all $l=1,\ldots,L-1$ for the
extreme binary distribution.

The same remains true for a bulk spin, which also has 
non-vanishing transverse magnetization only if it is weakly coupled
to the rest of the system (the trivial example being when both its
couplings to the left and to the right are exactly zero, which gives
the maximum value $m_1^x=1/2$). Thus the central spin in a chain of
length, say $2L-1$, has $m^x\sim{\cal O}(1)$ if and only if the
bond-configurations on both sides have surviving character, as it is
depicted in fig.\ \ref{fig4} for the extreme binary distribution.
Since the probability $P_{surv}(L/2)$ for a configuration of $L/2$
couplings to represent a surviving walk is $P_{surv}(L/2)\sim
L^{-1/2}$ it is
\be
m_l^x\sim\{P_{surv}(L/2)\}^2\sim L^{-1},\quad{\rm i.e.}\quad
x^x(XX)=1\;.
\ee
From this one obtains
\be
\eta^x(XX)=2\;,
\label{etaxxx}
\ee
as given in Table I. 

\begin{figure}
\epsfxsize=\columnwidth\epsfbox{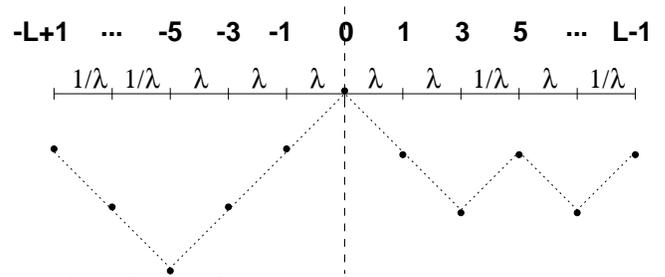}
\caption{Sketch of a bond configuration for a chain of length
  $2L-1$ that gives a non-vanishing transverse magnetization
  $m^x\sim{\cal O}(1)$ for the central (bulk) spin. The example is for
  the extreme binary distribution. Weak couplings ($J_{2i-1}=\lambda$)
  correspond to downward steps of the random walk on both sides of the
  central spin (here at $0$). Note that both, the right and the left
  half of the random walk have surviving character, i.e.\ do not cross
  the starting point.
}
\label{fig5}
\end{figure}

We verified the strong correlation between weak coupling and
non-vanishing transverse order parameter numerically in the following
way: We considered a chain with $L+1$ sites and the couplings at both
sides of the central spin were taken randomly from a distribution
called $SW$\cite{sw}, which represents those samples in the uniform distribution
which has a surface magnetization of $m_1^x(SW)>1/4$. (Thus cutting one of
the couplings to the central spin results a local magnetization greater
than $0.25$.) Then
we calculated numerically the order-parameter at the central spin and
its average value over the $SW$ configurations $[m_{L/2}^x]_{\rm
  sw}$ as given in Table I.

\bc
\begin{tabular}{|c|c|c|}
\hline
L&$2[m_1^x]_{\rm sw}$&$2[m_{L/2}^x]_{\rm sw}$\\
\hline
16&0.817&0.531\\
32&0.806&0.471\\
64&0.799&0.431\\
128&0.792&0.413\\
256&0.791&0.383\\
\hline
\end{tabular}
\ec
{\small TABLE II: Surface and bulk transverse order-parameters averaged 
over 50000 SW configurations for the uniform distribution.}
\bigskip

As seen in the Table the averaged surface order-parameter stays
constant for large values of $L$, whereas the bulk order-parameter
decreases very slowly, actually slower than any power. The data can be
fitted by $[m_{L/2}^x]_{\rm sw} \sim (\ln L)^{-\sigma}$, with $\sigma \approx 0.5$. Thus we
conclude that the numerical results confirm 
the exponents given in (\ref{etaxxx}), however there are
strong logarithmic corrections, 
which imply for the average transverse correlations
\be
[C^x(r)]_{\rm av} \sim r^{-2} \ln^{-1}(r)~~~{\rm XX-model}\;.
\label{CxXX}
\ee
These strong logarithmic corrections render the numerical calculation
of critical exponents very difficult \cite{girvin,stolzexx}. In
earlier numerical work using smaller finite systems disorder dependent
exponents were reported\cite{stolzexx}.  We believe that these
numerical results can be interpreted as effective, size-dependent
exponents and the asymptotic critical behavior is indeed described by
Eq. (\ref{CxXX}).

Note that our results in Table I satisfy the relation
$\eta^x(XX)=\eta^z(XX)$, both in the volume and at the surface, which
corresponds to Fisher's RG result\cite{fisherxx}. In this way we have
presented in independent justification of Fisher's RS phase picture,
where the average correlations are dominated by random singlets, so
that the distance between the pairs could be arbitrarily large.

\begin{figure}
\epsfxsize=\columnwidth\epsfbox{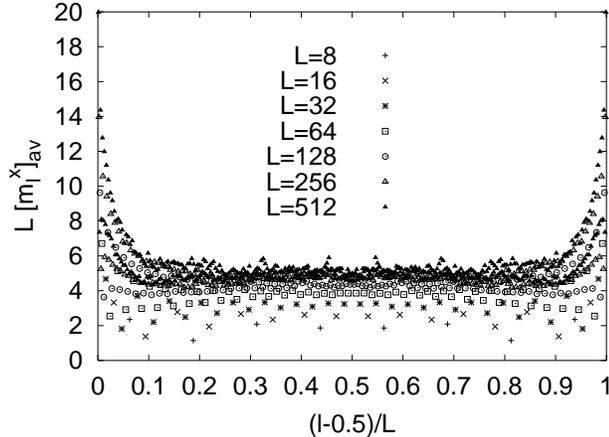}
\caption{Transverse order-parameter profile $[m_l^x]_{\rm av}$ for the
  $XX$ model at criticality for different system sizes calculated
  numerically with the fermion method using eq.
  (\protect{\ref{mx2}}). The data are for the uniform distribution,
  averaged over 50000 samples.}
\label{fig6}
\end{figure}

We checked numerically the above theoretical
predictions in the random $XX$ model.  In Fig. \ref{fig6} we present the
scaled $m_l^x$ profiles for the binary distribution for finite
systems up to $L=512$. The profiles have a broad plateau and the points
of $L^{x^x}m_l^x$ do not perfectly fall on one scaling curve due to strong
finite-size effects. Even system sizes as large as $L=512$ appear to 
be insufficient to get rid of such correction terms. Therefore we have
calculated the effective size-dependent $x^x(L)$ exponents by a two-point
fit. For this we have averaged the order-parameter in the middle of
the profile for $L/4<l<3L/4$ and compared this average values for
finite systems with $L/2$ and $L$ sites. As seen in Table III the
effective exponents are monotonously increasing with the size of the
system and they are not going to saturate, even for
$L=512$\cite{remark2}.

\bc
\begin{tabular}{|c|c|}
\hline
L&$x^x(L)$\\
\hline
16&0.635\\
32&0.677\\
64&0.730\\
128&0.823\\
256&0.872\\
512&0.910\\
\hline
\end{tabular}
\ec
{\small TABLE III: Effective bulk scaling dimension of the transverse 
order-parameter in the random $XX$ chain.}
\bigskip

From the data in Table III one can not make an accurate estimate about
the limiting value of $x^x(L)$, but it is clear that $x^x(L)$ grows at
least up to the theoretical limit $x^x=1$, although it could, in
principle, reach even a larger value. We note that similar observation
was made by Henelius and Girvin from the average $S^x$ correlation
function, where the effective $\eta^x$ exponents seem to grow over the
theoretically predicted value of $\eta^x=2$. (See Fig. 2 of
Ref\onlinecite{girvin}.)

\subsection{Autocorrelations}

\begin{figure}
\epsfxsize=\columnwidth\epsfbox{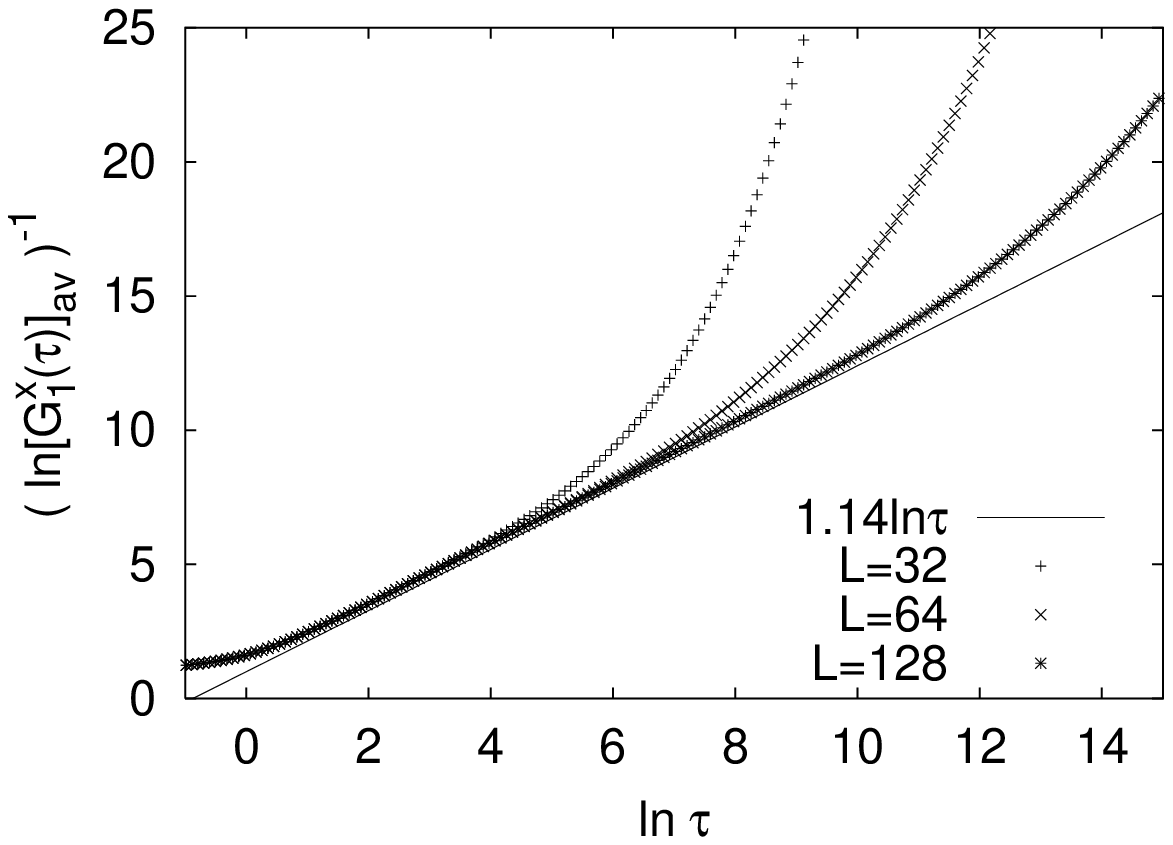}
\epsfxsize=\columnwidth\epsfbox{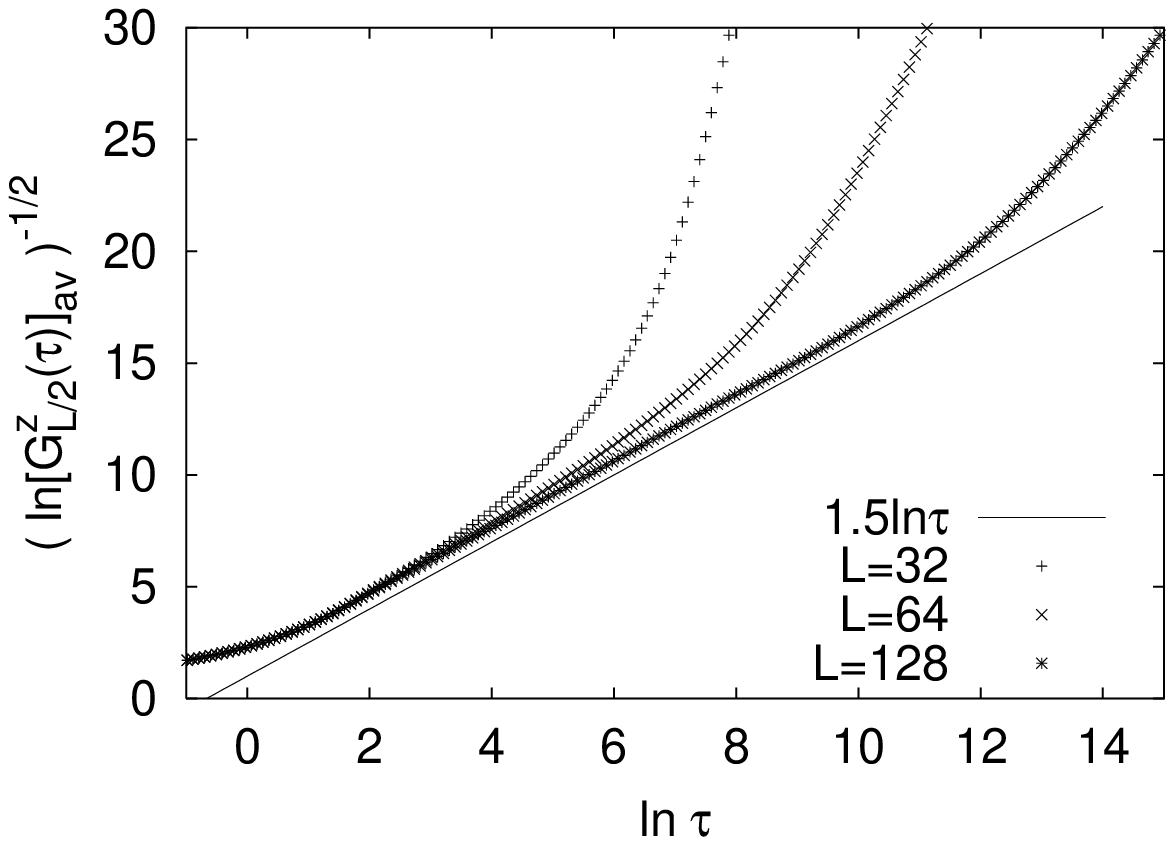}
\caption{
  Spin autocorrelation function
  $[G^{\mu}_l(\tau)]_{\rm av}$ for the 
  $XX$ model for $L=32,64$ and $128$ calculated numerically with the fermion method
  using eqs. (\protect{\ref{surfcorr}}) and (\protect{\ref{Glz}}). The data are for the binary
  distribution ($\lambda=4$), averaged over 50000 samples.
  a) (Top) shows $l=1$, the surface transverse autocorrelations, 
  b) (Bottom) shows $l=L/2$, the bulk longitudinal autocorrelations.}
\label{fig7}
\end{figure}

According to the scaling theory in Section III.C the decay of average
critical autocorrelations in random quantum spin chains is ultra-slow,
it takes place in logarithmic time-scales, as given in Eq.
(\ref{gx1}). Here we confront these predictions with the results of
numerical calculations.  We start with the surface autocorrelation
function $[G^x_1(\tau)]_{\rm av}$ for the $XX$ model, which is
calculated in the binary distribution ($\lambda=4$) on finite systems
up to $L=128$. As seen in fig.\ \ref{fig7} (top) the logarithmic
time-dependence is well satisfied and the decay exponent is found in
agreement with $\eta^x_1(XX)=1$ as given by the scaling result in Eq.
(\ref{gx1}).  For bulk spin critical autocorrelations we considered
$[G^z_{L/2}(\tau)]_{\rm av}$ for the $XX$ model.  Again the numerical
results in Fig.\ \ref{fig7} (bottom) are consistent with a logarithmic
decay with an exponent $\eta^z(XX)=2$, as given in Table I.

\begin{figure}
\epsfxsize=\columnwidth\epsfbox{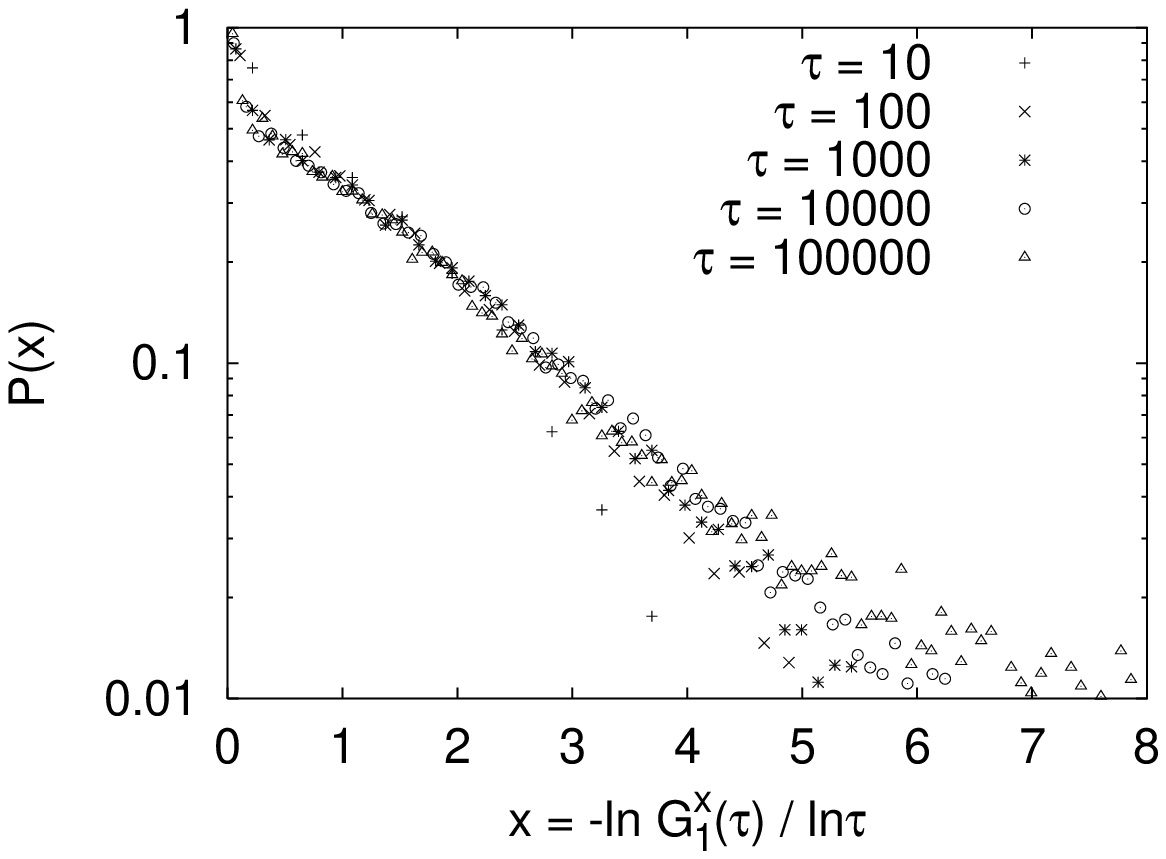}
\epsfxsize=\columnwidth\epsfbox{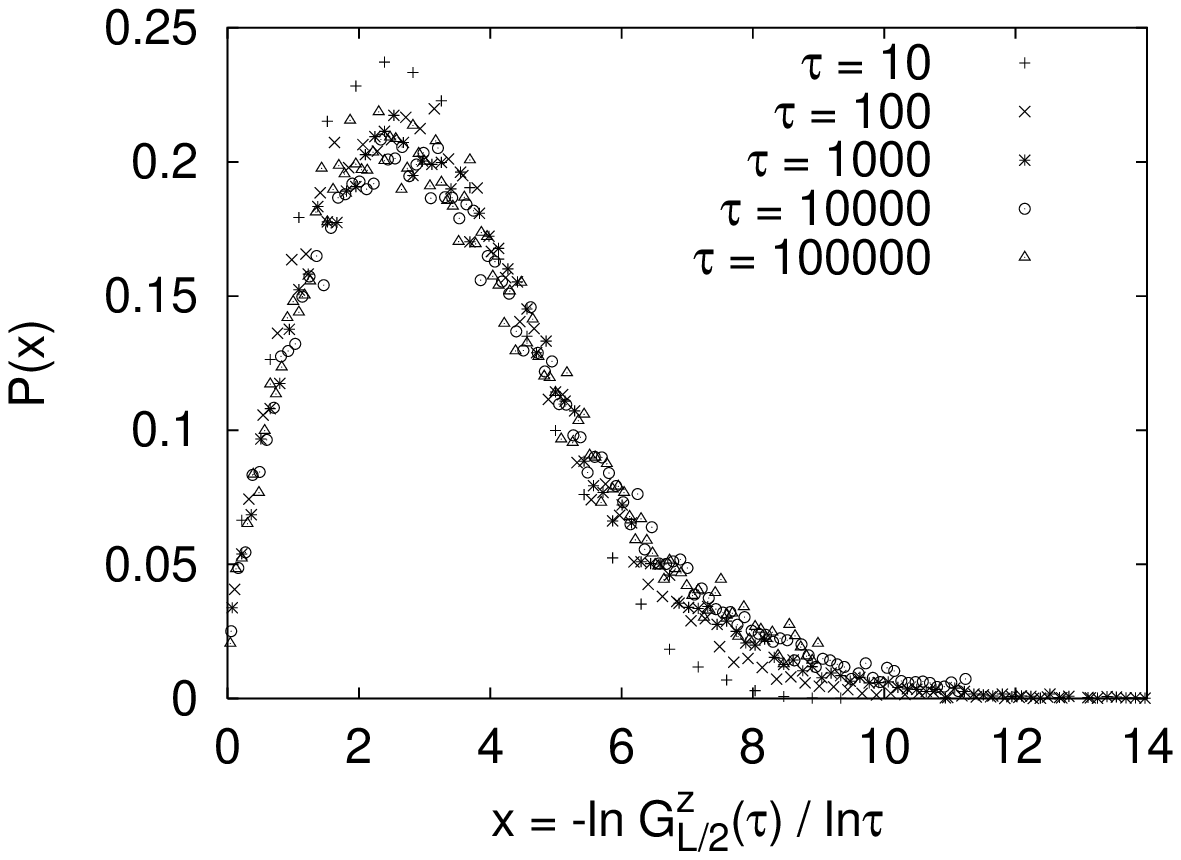}
\caption{Scaling plot of the probability distribution of the 
  autocorrelation function $G^{\mu}_l(\tau)$ for the $XX$-model
  for different values of $\tau$ at criticality ($L=128$). The data
  are for the uniform distribution averaged over 100000 samples.
  a) (Top) shows $l=1$, the surface transverse autocorrelations, 
  b) (Bottom) shows $l=L/2$, the bulk longitudinal autocorrelations.}
\label{fig8}
\end{figure}

Next we turn to study the {\it distribution} of critical
autocorrelations. As we have seen the average behavior is
logarithmically slow, but for typical samples, as described in
Appendix B, one expects a faster decay with a power-law
time-dependence. Then $G^{\mu}_l(\tau) \sim \tau^{-\gamma}$ and the
$\gamma$ exponent could vary from sample to sample. Such type of
``multi-scaling'' behavior of the autocorrelations has been recently
observed by Kisker and Young\cite{kiskeryoung} in the random quantum
Ising model. In Fig.\ \ref{fig8} we have numerically checked this
assumption for the critical autocorrelations $G^x_1(\tau)$ and
$G^z_{L/2}(\tau)$, respectively, of the random $XX$ chain, the average
behavior of those have been studied before. As seen in Fig.\ 
\ref{fig8} we have obtained indeed a good data collapse of the
probability distributions $P^{\mu}(\gamma)$ in terms of the scaling
variable $\gamma=-\ln G^{\mu}_l / \ln \tau$ for both type of
autocorrelations, but the scaling curve in the two cases are
different.

The average correlation function generally have a contributions from
the scaling function, $P^{\mu}(\gamma)$, but there could be also
non-scaling contributions, as found for the random quantum Ising chain
in Ref\cite{fisheryoung}. The scaling contribution is coming from the
small $\gamma$ part of the scaling function, which according to Fig.\ 
\ref{fig8} (top) for the autocorrelations $G^x_1(\tau)$ approaches a finite
value linearly, $P^x(\gamma) \sim A + B \gamma$.  Thus we have for the
average autocorrelations:
\beqn
[G_1^x(\tau)]_{\rm av} & = &\int_0^{\infty} P^x(\gamma) G_1^x(\tau) 
d \gamma \nonumber\\
& \sim & \int_0^{\infty} (A+B \gamma) \exp (-\gamma \ln \tau) 
d \gamma \nonumber\\
& \sim & A (\ln \tau)^{-1} + B (\ln \tau)^{-2}\;,
\label{g1xint}
\eeqn
in agreement with the scaling result in Eq. (\ref{gx1}) and with the
numerical result in Fig.\ \ref{fig8} (top) We note that the correction
to scaling contribution to the average autocorrelations in Eq.
(\ref{g1xint}) is also logarithmic.

For the critical autocorrelation $G^z_{L/2}(\tau)$ the scaling
function in Fig.\ \ref{fig8} (bottom) for small $\gamma$ approaches
linearly zero\cite{remark3} $P^z(\gamma) \sim \gamma$. Thus the
scaling contribution to the average autocorrelation, as evaluated
along the lines of Eq. (\ref{g1xint}), is $[G_1^x(\tau)]_{\rm av} \sim
(\ln \tau)^{-2}$, in agreement with the scaling result in Eq.
(\ref{gx1}).

\section{Griffiths phase}

Random quantum systems exhibit unusual off-critical properties: they
are gap-less in a extended region, $0<|\delta|<\delta_G$, as a result
of the so called Griffiths-McCoy singularities\cite{griffiths,mccoy}.
In this Griffiths phase the system is critical in the time direction,
although spatial correlations decay exponentially.

Quantitatively the basic information is contained in the distribution
of low energy excitations, $P(\epsilon)$, as given in Eq.
(\ref{edistr}).  With this the average autocorrelations can be
obtained as:
\be
[G(\tau)]_{av} \sim \int_0^{\infty} P(\epsilon) \exp(-\tau \epsilon) 
d \epsilon \sim \tau^{-1/z}\;,
\label{ggriff}
\ee
which is expected to hold for any component of the spin\cite{remark1}.
In this way we have recovered the scaling result in Eq.
(\ref{grdecay}).  In the Griffiths phase also some thermodynamic
quantities are singular, which are expressed as an integral of the
autocorrelation function. We mention the local susceptibility
$\chi^x_l$ at site $l$, which is defined through the local
order-parameter $m_l^x$ in Eq. (\ref{mx}) as
\be
\chi_l^x=\lim_{H_l^x \to 0}{\partial m_l^x \over \partial H_l^x}\;,
\ee
where $H_l^x$ is the strength of the local longitudinal field, which
enters the Hamiltonian in (\ref{hamilton}) via an additional term
$H_l^x S_l^x$. $\chi_l^x$ can be expressed as:
\be
\chi_l^x=2 \sum_{\langle n|}{|\langle n| S_l^x|0\rangle|^2\over E_n-E_0}\;,
\label{locsusc1}
\ee
thus its average value scales in finite systems as $\chi_l^x(L) \sim
L^{z-1}$, where we have used the scaling relation in Eq. (\ref{zL})
and the fact that the matrix-element in Eq. (\ref{locsusc1}) is $\sim
1/L$, since an SCD can be localized at any site of the chain. For a
small finite temperature $T$ we can use the scaling relations $T \sim
\epsilon \sim L^{-z}$ and we have for the singular behavior:
\be
[\chi_l^x(T)]_{\rm av} \sim T^{-1+1/z}\;.
\ee
To estimate the temperature dependence of the average specific heat,
$[C(T)]_{\rm av}$, we calculate first the average excitation energy
per SCD with $P(\epsilon)$ in Eq. (\ref{edistr}) as $\int \epsilon
P(\epsilon) d \epsilon \sim \epsilon^{1/z+1}$, which is proportional
to the thermal excess energy per spin $\sim T^{1/z+1}$, from which we
obtain:
\be
[C(T)]_{\rm av} \sim T^{1/z}\;.
\ee
We note that several other physical quantities are singular in the
Griffiths phase (non-linear susceptibility, higher excitations, etc)
and the corresponding singularities are expected to be related to the
dynamical exponent $z$. For a detailed study of this subject in the
random quantum Ising model see Ref.\onlinecite{ijr}.

In the following we calculate the exact value of the dynamical
exponent using the same strategy as for the random quantum Ising model
in Ref\onlinecite{diffusion,itr}.  Our basic observation is the fact
that the eigenvalue problem of the $T_{\sigma}$ (or $T_{\tau}$) matrix
can be mapped through an uniter transformation to a Fokker-Planck
operator, which appear in the Master equation of a Sinai diffusion,
i.e. random walk in a random environment\cite{sinai}. The transition
probabilities of the latter problem are then expressed with the
coupling constants of the spin model. The Griffiths phase of the spin
model corresponds to the anomalous diffusion region of the Sinai
walk and from the exact results about the scaling form of the energy
scales in this problem one obtains for the dynamical exponent of the
$XY$ model:
\be
\left[ \left( J^x \over J^y \right)^{1/z} \right]_{\rm av}=1\;,
\label{zexact}
\ee
whereas for the $XX$ model the result follows with the correspondences
in Eq.(\ref{xyxx}).  For the binary distribution in Eq. (\ref{binary})
the Griffiths phase is for $1 < J^y_0 < \lambda$ and $z$ is given by:
\be
\left(J_0^y \right)^{1/z}={\rm cosh}\left( {\ln \lambda \over z} \right)\;.
\ee
For the uniform distribution
\be
z \ln\left(1-z^{-2}\right) = - \ln J^y_0\;,
\ee
and the Griffiths phase extends to $1< J^y_0 < \infty$.

Next we are going to study numerically the Griffiths phase and to
verify some of the scaling results described above. In this respect we
shall not consider those quantities which have an equivalent
counterpart in the random quantum Ising model (distribution of energy
gaps, local susceptibility, specific heat, etc), since that model has
already been thoroughly investigated
numerically\cite{youngrieger,young,bigpaper,ijr}. The autocorrelation
functions, however, are different in the two models and we are going
to study those in the following.

\begin{figure}
\epsfxsize=\columnwidth\epsfbox{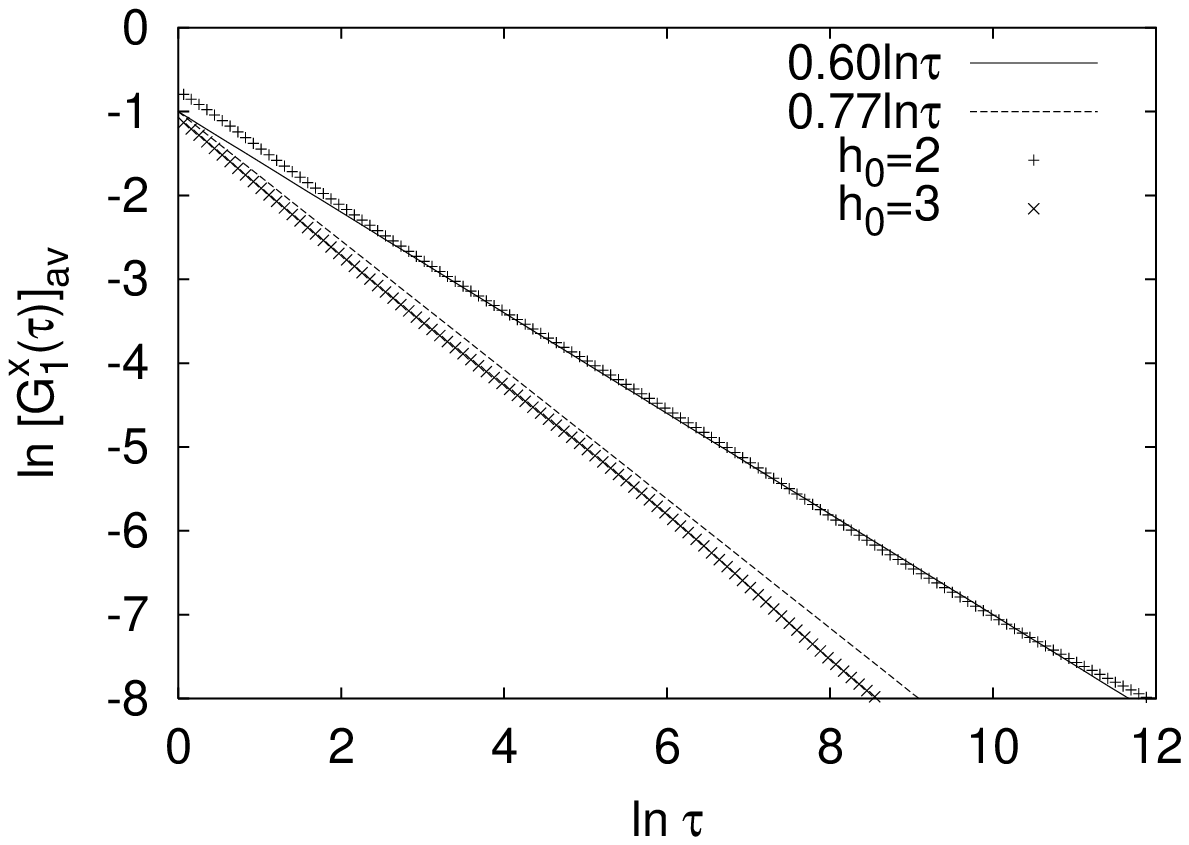}
\epsfxsize=\columnwidth\epsfbox{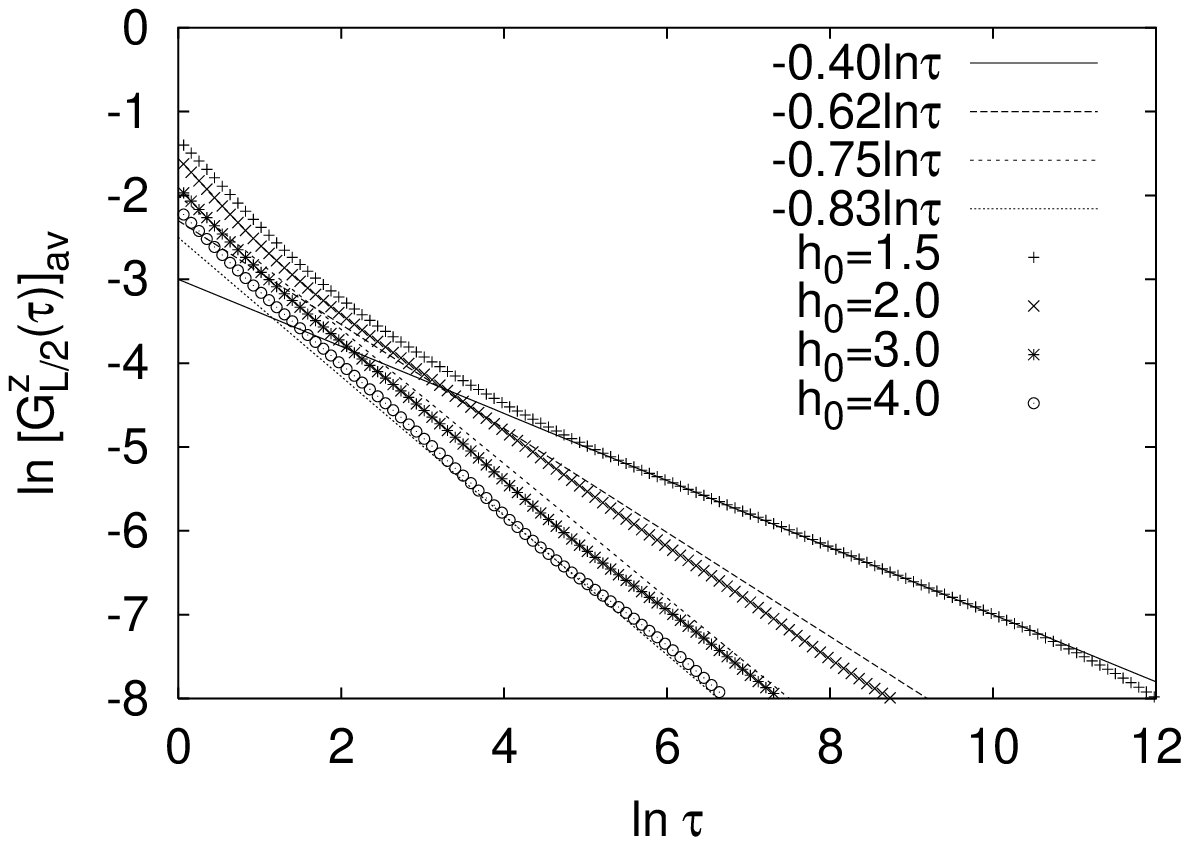}
\caption{
  The average surface (top) and bulk (bottom)
  autocorrelation function $[G^{\mu}_{L/2;1}(\tau)]_{\rm av}$ of the $XX$
  model in the Griffiths-phase for various values of $h_0$. The
  straight lines have a slope of $1/z(h_0)$, where the dynamical
  exponent $z(h_0)$ agrees well the exact value determined via the
  formula (\ref{zexact}). The data are for the uniform distribution
  averaged over 50000 samples of size $L=128$.}
\label{fig9}
\end{figure}

The average bulk longitudinal autocorrelation function
$[G^z_{L/2}(\tau)]_{\rm av}$ of the $XX$ model is shown in Fig.\ 
\ref{fig9} in a log-log plot at different points of the Griffiths
phase. The asymptotic behavior in Eq. (\ref{ggriff}) is well satisfied
and the dynamical exponents obtained from the slope of the curves are
in good agreement with the analytical results in Eq. (\ref{zexact}).
Similar conclusion can be drawn from the average surface transverse
autocorrelations, $[G^x_{1}(\tau)]_{\rm av}$, as shown in Fig.\ 
\ref{fig9}.

\begin{figure}
\epsfxsize=\columnwidth\epsfbox{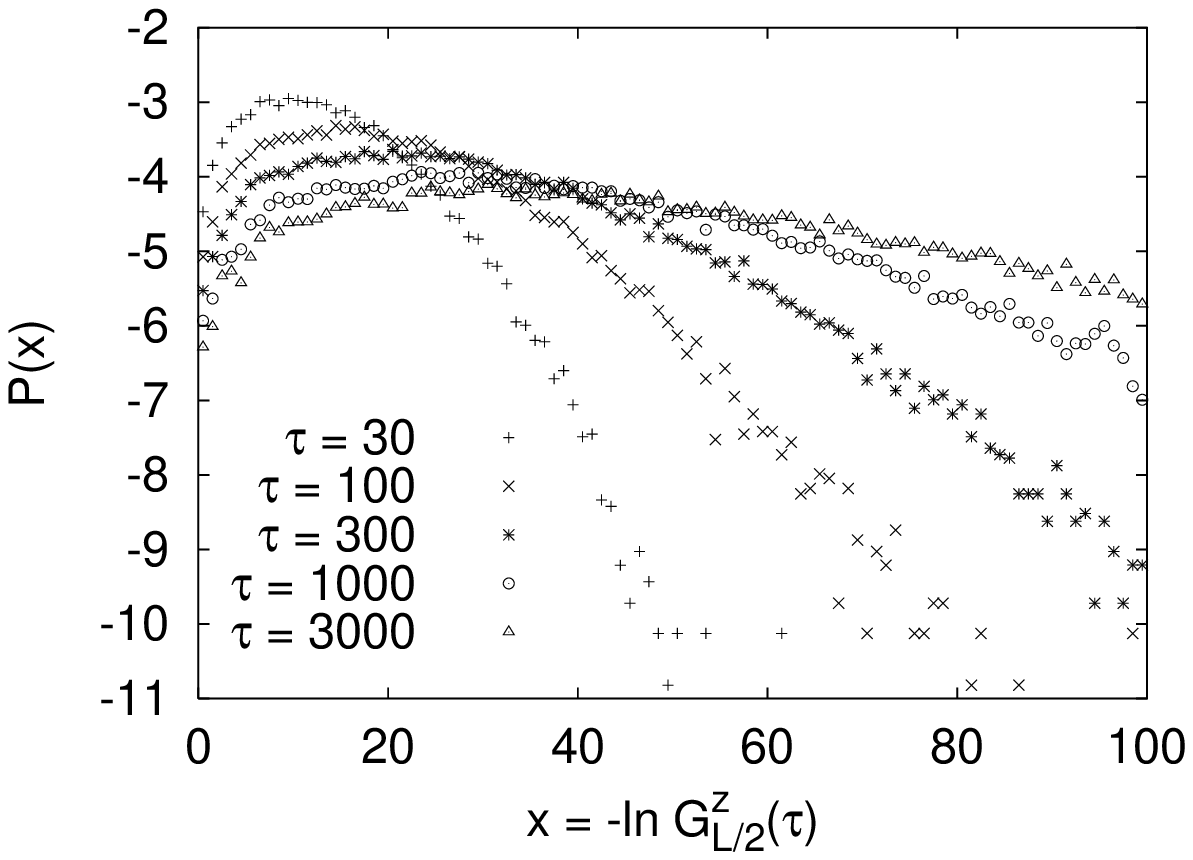}
\epsfxsize=\columnwidth\epsfbox{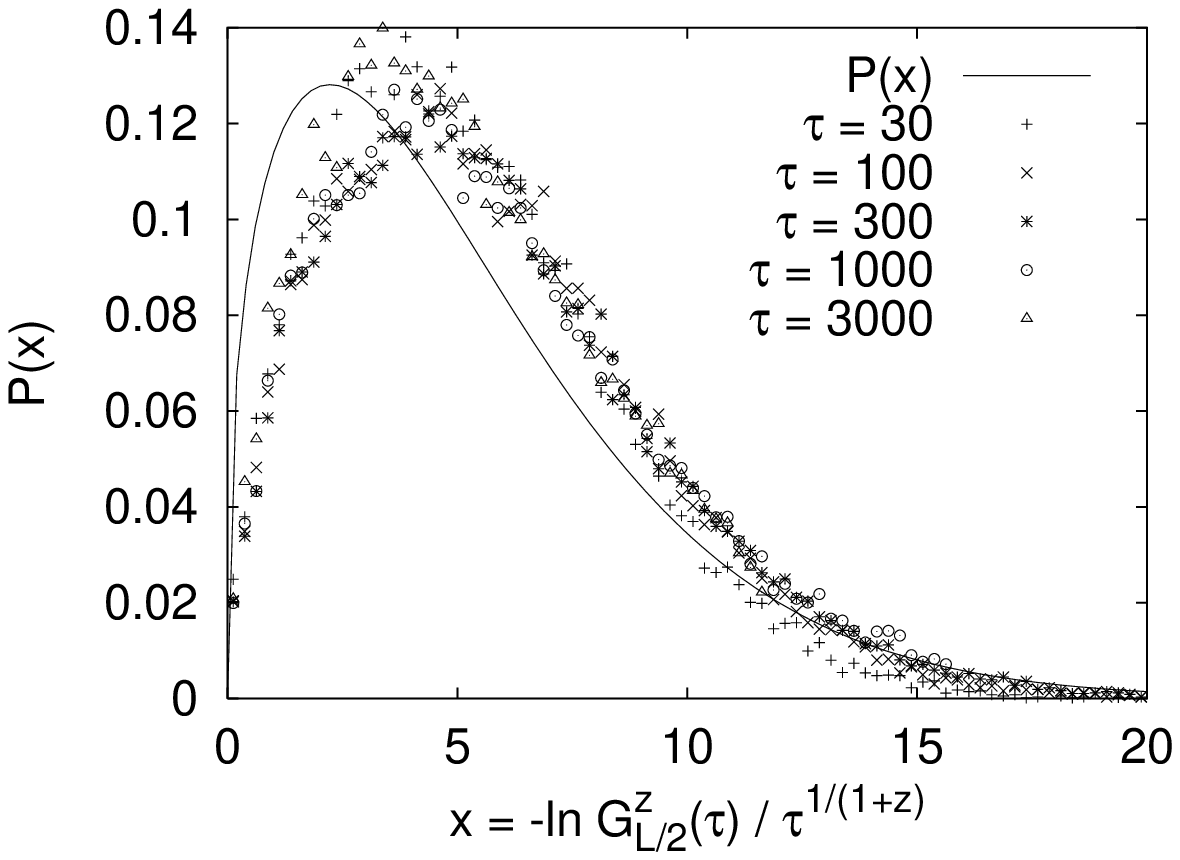}
\caption{
  (Top): Probability distribution of the bulk longitudinal
  autocorrelation function of the $XX$ model in the Griffiths phase
  for $h_0=1.5$.  The data are for the uniform distribution from
  100000 samples of size $L=128$.\\
  (Bottom): Scaling plot of the data in the top figure.  The scaling
  variable $[\ln G(\tau)] / \tau^{1/(z+1)}$ contains the dynamical
  exponent $z(h_0)$ known from the formula (\ref{zexact}). The full
  curve is the theoretical prediction in (\protect{\ref{young_pred}})
  using the exact value of $z(h_0=1.5)=2.659$ and a fit-parameter
  $c=0.22$.}
\label{fig10}
\end{figure}

Next we study the distribution of the autocorrelation functions. In
Fig.\ \ref{fig10} the distribution of the bulk longitudinal
autocorrelation function of the $XX$ model is given at different times
$\tau$. As argued in the Appendix the typical autocorrelations are of
a stretched exponential form
\be
G(\tau) \sim \exp\left( -{\rm const}~\tau^{1/(1+z)} \right)\;,
\ee
thus the relevant scaling variable is
\be
\alpha=-{\ln G(\tau) \over \tau^{1/(z+1)}}\;.
\label{gdistr}
\ee
Using this scaling argument we obtained a good data collapse of the
points of the distribution function as shown in Fig.\ \ref{fig10} . We
note that for the random quantum Ising model Young\cite{young} has
also derived the scaling function from phenomenological arguments,
\be
P(x)=c(cx)^{1/z}\,\exp\left(-\frac{z}{1+z} (cx)^{1+1/z}\right) 
\label{young_pred}
\ee
which is also presented in Fig. \ref{fig10}. One can see considerable
differences between the numerical and theoretical curves. Similar
tendencies have been noticed for the random quantum Ising model in
Ref\onlinecite{young}.  The discrepancies are probably due to strong
correction to scaling or finite size effects. These corrections,
however does not affect the scaling form in Eq. (\ref{gdistr}).

\section{Discussion}

In this section we first discuss the possible extension of our results
to random $XXZ$ chains and to higher dimensional systems, then we
conclude with a brief summary of our findings.

\subsection{Random $XXZ$ chains}

The more general $XXZ$ (or $XYZ$) Heisenberg spin chain, where the
Hamiltonian in Eq. (\ref{hamilton}) contains an additional interaction
term of the form
\be H_Z=\sum_{l=1}^{L-1} J_l^z S_l^z S_{l+1}^z\;, \label{HZ} 
\ee
can be treated perturbatively when $J^z_l \ll J^x,J^y$. Here we
consider the $XXZ$ chain, with $J^z_l=\lambda J_l$ and $\lambda \ll 1$.
To see the scaling behavior of the energy gap we express the small
perturbation, $H_Z$, in terms of two decoupled TIM-s (see Appendix) as 
\be 
H_Z=-{1 \over 4}\sum_{i=1}^{L/2-1} \left(J_{2i-1}^z \sigma_i^z
\tau_i^z + J_{2i}^z \sigma_i^x \sigma_{i+1}^x \tau_i^x \tau_{i+1}^x
\right)\;, 
\label{HZ1} 
\ee 
whose expectation value in the unperturbed ground state is just the
product of the local energy-densities of the TIM-s. The perturbative
correction to the gap, $\Delta_z(L)=\langle 1|H_z|1 \rangle - \langle
0|H_z|0 \rangle$, evaluated with the states of the unperturbed
Hamiltonian, $\langle 0|$ and $\langle 1|$, is proportional to the gap
of one of the random TIM's, $\epsilon_{\sigma}(L)$. At the critical
point $\epsilon_{\sigma}(L) \sim \exp(-{\rm const}\cdot
L^{1/2})$\cite{bigpaper}, which is the same scaling form as for the
$XY$ model in Eq.(\ref{scener}). Thus the scaling relation in Eq.
(\ref{e1.1}) is valid also for the XXZ chain. In the Griffiths-phase one
has again $\epsilon_{\sigma}(L)\sim L^{-z}$, and the dynamical
exponent $z$, has the same value as in Eq. (\ref{zexact}). Thus we
arrive at the conclusion that also in the Griffiths phase the
corresponding scaling relation in Eq. (\ref{z}) is valid in the same
form for the random $XXZ$ chain, at least for small $J^z$ couplings.

Next we study the asymptotic properties of average critical correlations
in the random $XXZ$ model through the scaling behavior of the local
order-parameters. As we argued in Section III these quantities are
related to the fraction of rare events, $P^{\mu}$, and here we are going
to investigate the influence of the perturbation  $H_Z$ to $P^{\mu}$. We
start with the surface transverse order-parameter $m_1^x$ and recall
that it is maximal, i.e. $m_1^x=1/2$, if the surface spin is
disconnected in the $XY$-plane, i.e. $J^x_1=J^y_1=0$. Evidently the
value of $m_1^x=1/2$ does not change for any finite value of the
coupling $J^z_1$. Now consider the infinite-randomness fixed point of
the $XX$ chain with the extreme binary distribution, where a rare event
is represented by couplings with a surviving random walk configuration
and with $m_1^x=O(1)$. Roughly speaking, a rare event is formally
equivalent to a situation, in which there is a very weak surface
coupling of $J_1=O(L^{-1/2})$, where $L$ is the system size.  Then
switching on homogeneous and finite couplings $J^z$ the lowest
excitation of the chain stays localized at the surface, since the shape
of the wave-function does not change significantly in first order
perturbation theory. Consequently the surface order-parameter is still
$m_1^x=O(1)$, and the sample is a rare event for the $XXZ$ chain, too.
For small random couplings $J_l^z$ the accumulated fluctuations in
$J_l^z$ are divergent as $\sim L^{1/2}$, however these are still
negligible compared with the fluctuations in the transverse couplings.
Thus the rare events of the $XX$ chain are identical with those
appearing in the $XXZ$ chain for small values of the random longitudinal
couplings. As a consequence the critical end-to-end average correlations
decay with the same exponent as given in Table I. Since the rare events
for other local order-parameters are also connected to SCD-s with
localized wave-functions the stability of the infinite-randomness fixed
point holds for the other critical correlations, too. Actually, it seems
to be plausible that the attracting region of the $XX$ fixed point
extends up to $[\ln J^x]_{\rm av} > [\ln J^z]_{\rm av}$, i.e. where the
average transverse couplings are larger than the longitudinal ones, thus
up to the random $XXX$ fixed point, in agreement with Fisher's
conjecture\cite{fisherxx}.

\subsection{Higher dimensions}

In one dimension the topology is special since there is only a single
path between two points, whereas in higher dimensional lattices one
has several distinct paths connecting two points. This topological
difference is essential when random $XX$ magnets are considered in
higher dimensions. Let us consider again the surface transverse
order-parameter and construct a rare event in the extreme binary
distribution. For this purpose the surface spin should be extremely
weakly coupled to the bulk of the system. Thus considering any
non-self-crossing path from the spin to the volume one should have a
surviving random walk configuration in the couplings. In higher
dimensions the number of such paths grows exponentially with the size
of the system, $L$, thus the fraction of rare events, which is related
to the length as a power in one dimensions, becomes exponentially
small in higher dimensions. Consequently the infinite-randomness fixed
point picture is not applicable here and one concludes that the
critical properties of higher dimensional $XX$ and Heisenberg
antiferromagnets are controlled by conventional random fixed points.
This result is also in agreement with numerical RG calculations in
2d\cite{RG}. We note that in contrast to random Heisenberg
antiferromagnets the random ferromagnetic quantum Ising models in
higher dimensions are still controlled by infinite-randomness fixed
points\cite{fm2d,RG}.

\subsection{Summary}

Quantum spin chains in the presence of quenched disorder show unusual
critical properties, which are controlled by the infinite-randomness
fixed point. A common feature of these systems is that various physical
properties, especially those related to local order-parameters and
correlation functions are not self-averaging and their average
behavior is determined by the rare events (or rare regions), which
give the dominant contribution, although their fraction is vanishing
in the thermodynamic limit. In this paper we have performed a detailed
study of the scaling behavior of rare events appearing in the random
$XY$ and $XX$ chains.  We identified the rare events as strongly
coupled domains, where the coupling distribution follows some
surviving random walk character. From the scaling properties of the
rare events we have identified the complete set of critical decay
exponents and found exact results about the correlation length
exponent and the scaling anisotropy.

Another new aspect of our work was the study of dynamical correlations. We
have obtained the asymptotic behavior of the average autocorrelation
function and determined the scaling form of the distribution of
autocorrelations. In the off-critical regime we investigated the
singular physical quantities in the Griffiths phase. In particular we
have obtained exact expression for the dynamical exponent $z$, which is a
continuous function of the quantum control-parameter and the
singularities of all physical quantities can be related to its value.

\section*{Acknowledgements}
This work has been partially performed during our
visits in K\"oln and Budapest, respectively. F.\ I.'s work has been
supported by the Hungarian National Research Fund under grant No OTKA
TO23642, TO25139, MO28418 and by the Ministery of Education under
grant No. FKFP 0596/1999.  H.\ R. was supported by the Deutsche
Forschungsgemeinschaft (DFG). We thank L. Turban for helpful comments
on the manuscript.

\appendix

\section*{Mapping to decoupled Ising quantum chains}

We start here with the observation in Section IIB that the eigenvalue
matrix $T$ in Eq. (\ref{tmatrix}) can be represented as a direct
product $T=T_{\sigma} \bigotimes T_{\tau}$.  The trigonal matrices
$T_{\sigma}$ , $T_{\tau}$ of size $L \times L$ represent transfer
matrices of directed walks, which are in one-to-one correspondence
with Ising chains in transverse field\cite{igloiturban96} defined by
the Hamiltonians:
\beqn
H_{\sigma}&=&-{1 \over 4} \sum_{i=1}^{L/2-1} J^x_{2i} \sigma_i^x \sigma_{i+1}^x
-{1 \over 4}\sum_i^{L/2} J^y_{2i-1} \sigma_i^z\nonumber\\
H_{\tau}&=&-{1 \over 4}\sum_{i=1}^{L/2-1} J^y_{2i} \tau_i^x \tau_{i+1}^x
-{1 \over 4}\sum_i^{L/2} J^x_{2i-1} \tau_i^z\;.
\label{Hst}
\eeqn
Here the $\sigma^{x,z}_i$ and $\tau^{x,z}_i$ are two sets of Pauli
matrices at site $i$ and there are free boundary conditions for both
chains. We can then write $H_{XY}=H_{\sigma}+H_{\tau}$.  Note the
symmetry $\sigma_i^{x,z} \leftrightarrow \tau_i^{x,z}$ and $J_l^x
\leftrightarrow J_l^y$, thus anisotropy in the $XY$ model has
different effects in the two Ising chains.

One can easily find the transformational relations between the $XY$
and Ising variables:
\beqn
\sigma_i^x&=&\prod_{j=1}^{2i-1} \left( 2S_j^x \right),~~~\sigma_i^z=4 S^y_{2i-1} S^y_{2i}
\nonumber\\
\tau_i^x&=&\prod_{j=1}^{2i-1} \left( 2S_j^y \right),~~~\tau_i^z=4 S^x_{2i-1} S^x_{2i}\;,
\label{oprel}
\eeqn
whereas the inverse relations are the following:
\beqn
2 S_{2i-1}^x&=&\sigma_i^x \prod_{j=1}^{i-1} \tau_j^z ,~~~2 S_{2i}^x=\sigma_i^x\prod_{j=1}^{i} \tau_j^z
\nonumber\\
2 S_{2i-1}^y&=&\tau_i^x \prod_{j=1}^{i-1} \sigma_j^z ,~~~2 S_{2i}^y=\tau_i^x\prod_{j=1}^{i} \sigma_j^z\;.
\label{invrel}
\eeqn
We note that a relation between the $XY$ model and two decoupled Ising
quantum chains in the thermodynamic limit is known for some
time\cite{peschel,fisherxx}, here we have extended this relation for
finite chains with the appropriate boundary conditions. These are
essential to map local order-parameters and end-to-end correlation
functions.

End-to-end correlations are related as
\be
\langle S_1^x S_L^x \rangle={1 \over 4} \langle \sigma_1^x \sigma_{L/2}^x \rangle \langle
\prod_{i=1}^{L/2} \tau_i^z \rangle={1 \over 4} \langle \sigma_1^x \sigma_{L/2}^x \rangle\;,
\ee
since in the ground state $\langle \prod_{i=1}^{L/2} \tau_i^z \rangle=1$. Similarly
\be
\langle S_1^y S_L^y \rangle={1 \over 4} \langle \tau_1^x \tau_{L/2}^x \rangle\;,
\ee
thus the end-to-end correlations in the two models are in identical
form. As a consequence the corresponding decay exponent in the random
models, $\eta_1^x$ in Table I is the same in the two systems and the
same conclusion holds also for the correlation length exponent, $\nu$
in Eq. (\ref{nu2}). These results are also independent of the type of
correlation of the disorder, thus are valid both for the $XY$ and $XX$
models.

Correlations between two spins at general positions $2l$ and $2l+2r$
are related as
\be
\langle S_{2l}^x S_{2l+2r}^x \rangle={1 \over 4} \langle \sigma_l^x \sigma_{l+r}^x \rangle \langle
\prod_{i=1}^{r} \tau_{l+i}^z \rangle\;.
\label{corrtr}
\ee
The second factor in the r.h.s., $\langle \prod_{i=1}^{r} \tau_{l+i}^z
\rangle$, defines a string-like order-parameter\cite{girvin} what can
be expressed in a simpler form in terms of the dual Ising variables
$\tilde{\tau}^x_{i+1/2}$, which are defined on the bonds of the
original Ising chain as
\beqn
\tilde{\tau}^z_{i+1/2}&=&\tau_i^x \tau_{i+1}^x\nonumber\\
\tau_i^z&=&\tilde{\tau}^x_{i-1/2} \tilde{\tau}^x_{i+1/2}\;.
\eeqn
Under the duality transformation fields and couplings are exchanged,
therefore the vanishing bonds at the two ends of an open chain are
transformed to vanishing fields, thus the dual chain has two end spins
fixed to the same state. So we obtain for the correlations in Eq.
(\ref{corrtr})
\be
\langle S_{2l}^x S_{2l+2r}^x \rangle={1 \over 4} \langle \sigma_l^x \sigma_{l+r}^x \rangle \langle
\tilde{\tau}_{l+1/2}^x \tilde{\tau}_{l+r+1/2}^x \rangle^{++}\;,
\label{corrtr1}
\ee
where the superscript $^{++}$ denotes fixed-spin boundary condition.
For non-surface points the average value of the correlation function
in Eq. (\ref{corrtr1}) depends on the type of disorder correlations.
For the $XY$ model, where the disorder is uncorrelated the two factors
in Eq. (\ref{corrtr1}) can be averaged separately, whereas this is not
possible for the $XX$ model.  We treated this point in Sections
IV.B.2.

\section*{Distribution of autocorrelation functions}

The autocorrelation functions is represented by the general form:
\be
G(\tau)=\sum_k |M_k|^2 \exp(-\tau \Delta E_k)\;
\label{gtau}
\ee
where the dominant contributions to the sum in Eq. (\ref{gtau}) are
from SCD-s, which are localized at some distance $l$ from the spin and
have a very small excitation energy, $\Delta E(l)$.  The scaling form
of $\Delta E(l)$ follows from the considerations in Section III.B and
one obtains from Eqs. (\ref{scener}) and (\ref{zL})
\be
\Delta E(l) \sim \cases{ \epsilon_0 \exp(-A l^{1/2}),\delta=0\cr
                \epsilon_0 l^{-z(\delta)},\delta<0\cr}\;,
\label{epsl}
\ee
at the critical point and in the Griffiths phase, respectively, where
$\epsilon_0$ denotes the energy scale. Thus the larger the distance
from the spin the larger the probability to have an SCD with a very
small energy.  For the matrix-element, $|M(l)|^2$, the tendency is the
opposite since the overlap with the wave-function of the SCD is
(exponentially) decreasing with the distance. The corresponding
scaling form can be read from the typical behavior of the surface
order-parameter as given below and above Eq. (\ref{nutyp}) as
\be
|M(l)|^2 \sim \cases{\exp(-B l^{1/2}),~~~\delta=0\cr
                \exp(-l/\xi_{\rm typ}),~~~\delta<0\cr}\;.
\label{ml}
\ee
Then $G(\tau)$ in Eq. (\ref{gtau}) can be approximated by a sum which
runs over SCD-s localized at different distances, $l$, and this sum is
dominated by the largest term with $l=l_0$:
\be
G(\tau) \sim |M(l_0)|^2 \exp(-\tau \Delta E(l_0))\;.
\label{gtau0}
\ee
Using the scaling forms in Eqs. (\ref{epsl}) and (\ref{ml}) one gets
following result.

At the critical point the characteristic distance is $l_0=[\ln(\tau
\epsilon_0 A / B)/A]^2$ and the typical autocorrelation function
decays as a power:
\be
G(\tau) \sim \tau^{-B/A},~~~\delta=0\;,
\ee
Thus the relevant scaling variable of the problem is
\be
\gamma=-{\ln G(\tau) \over \ln \tau},~~~\delta=0\;.
\ee
In the Griffiths phase the characteristic distance has a power-law
$\tau$ dependence, $l_0=\xi_{\rm typ} (\tau \epsilon_0 z)^{1/(z+1)}$,
which is however different from the average scaling form in Eq.
(\ref{z}).  The typical autocorrelations now are in a stretched
exponential form:
\be
G(\tau) \sim \exp\left[-(\tau \epsilon_0 z)^{1/(z+1)} \left( 1 +{1 \over z} \right) \right],
~~~\delta \ne 0\;,
\ee
and the relevant scaling variable is given in Eq. (\ref{gdistr}).

\end{multicols}

\begin{references}
\bibitem{remark2} It is not feasible to increase the size of the
        system further, primary not due to computational demand, but
        due to inaccuracies in the numerical routines, even in 64-bit
        precision. The origin of this numerical difficulty are those
        samples with an extremely small excitation energy.
\bibitem{remark3}
        For a finite system the scaling function approaches a finite limiting value of
        $P^z(\gamma=0,L) \sim L^{-1/2}$, which is checked numerically.
\bibitem{remark1} For correlations of such spin components which
        exhibit long range order one considers the connected
        autocorrelation function and in Eq. (\ref{edistr})) the
        distribution is related to the second gap.
\end{references}
\end{document}